\pdfoutput=1
\RequirePackage{ifpdf}
\ifpdf 
\documentclass[pdftex]{sigma}
\else
\documentclass{sigma}
\fi

\numberwithin{equation}{section}

\begin{document}


\newcommand{\arXivNumber}{1705.06424}

\renewcommand{\thefootnote}{}

\renewcommand{\PaperNumber}{080}

\FirstPageHeading

\ShortArticleName{Integrable Deformations of Sine-Liouville Conformal Field Theory and Duality}

\ArticleName{Integrable Deformations of Sine-Liouville\\ Conformal Field Theory and Duality\footnote{This paper is a~contribution to the Special Issue on Recent Advances in Quantum Integrable Systems. The full collection is available at \href{http://www.emis.de/journals/SIGMA/RAQIS2016.html}{http://www.emis.de/journals/SIGMA/RAQIS2016.html}}}

\Author{Vladimir~A.~FATEEV~$^{\dag\ddag}$}

\AuthorNameForHeading{V.A.~Fateev}

\Address{$^\dag$~Laboratoire Charles Coulomb UMR 5221 CNRS-UM2, Universit\'e de Montpellier,\\
\hphantom{$^\dag$}~34095 Montpellier, France}
\EmailD{\href{mailto:Vladimir.Fateev@univ-montp2.fr}{Vladimir.Fateev@univ-montp2.fr}}

\Address{$^\ddag$~Landau Institute for Theoretical Physics, 142432 Chernogolovka, Russia}

\ArticleDates{Received April 24, 2017, in f\/inal form October 03, 2017; Published online October 13, 2017}

\Abstract{We study integrable deformations of sine-Liouville conformal f\/ield theory. Eve\-ry integrable perturbation of this model is related to the series of quantum integrals of motion (hierarchy). We construct the factorized scattering matrices for dif\/ferent integrable perturbed conformal f\/ield theories. The perturbation theory, Bethe ansatz technique, renormalization group and methods of perturbed conformal f\/ield theory are applied to show that all integrable deformations of sine-Liouville model possess non-trivial duality properties.}


\Classification{16T25; 17B68; 83C47}

\renewcommand{\thefootnote}{\arabic{footnote}}
\setcounter{footnote}{0}

\section{Introduction}

Duality plays an important role in the analysis of statistical, quantum f\/ield and string theory systems. Usually it maps a weak coupling region of one theory to the strong coupling region of the other and makes it possible to use perturbative, semiclassical and renormalization group methods in dif\/ferent regions of the coupling constant. For example, the well known duality between sine-Gordon and massive Thirring models \cite{C,M} together with integrability plays an important role for the justif\/ication of exact scattering matrix \cite{SGZ} in these theories. Another well known example of the duality in two-dimensional integrable systems is the weak-strong coupling f\/low from af\/f\/ine Toda theories to the same theories with dual af\/f\/ine Lie algebra \cite{AFT,AFT+,AFT++}. The phenomenon of electric-magnetic duality in four-dimensional $N=4$ supersymmetric gauge theories conjectured in \cite{NO, MO} and developed for $N=2$ theories in~\cite{SW} (and in many subsequent papers) opens the possibility for the non-perturbative analysis of the spectrum and phase structure in supersymmetric gauge f\/ield theories. Recently discovered remarkable f\/ield/string duality \cite{ADSCFT+++,ADSCFT+,ADSCFT,ADSCFT++} leads to the unif\/ication of the ideas and methods for the analysis of these seemingly dif\/ferent quantum systems.

Known for many years the phenomenon of duality in quantum f\/ield theory still looks rather mysterious and needs further analysis. This analysis essentially simplif\/ies for two-dimensional integrable relativistic systems. These theories besides the Lagrangian formulation possess also unambiguous def\/inition in terms of factorized scattering theory, which contains all information about of\/f-mass-shell data of quantum theory. These data permit one to use non-perturbative methods for the calculation of observables in integrable f\/ield theories. The comparison of the observables calculated from the scattering data and from the perturbative, semiclassical or renormalization group analysis based on the Lagrangian formulation makes it possible in some cases to justify the existence of two dif\/ferent (dual) representation for the Lagrangian description of quantum theory.

Two particle factorized scattering matrix is rather rigid object. It is constrained by the global symmetries, factorization equation and unitarity and crossing symmetry relations. After resolution of these equations the scattering matrix $S$ can contain one (or more) free parameter. At some value of this parameter $\lambda=\lambda_{0}$ the scattering matrix $S( \lambda_{0})$ becomes identity matrix and possesses the regular expansion at $( \lambda -\lambda _{0}) $ near this point. In many cases this expansion can be associated with perturbative expansion of some Lagrangian theory with parameter $b$ near some free point. In some cases $S(\lambda)$ contains other point $\lambda _{1}$ where $S(\lambda _{1})$ is the identity matrix and admits the regular expansion in $( \lambda -\lambda _{1})$. If this expansion can be associated with some perturbative expansion with other local Lagrangian and small coupling $\gamma =\gamma (b)$, then two dif\/ferent Lagrangians describe the same theory, which possesses two dif\/ferent (dual) perturbative regimes.

More interesting situation occurs when $S(\lambda) $ has the regular expansion in $(\lambda -\lambda_{0})$ which is in perfect agreement with perturbative expansion in~$b$ of some f\/ield theory with local action~$\mathcal{A}(b)$, but at the point~$\lambda _{1}$ the $S$-matrix tends to some ``rational'' scattering matrix corresponding to the $S$-matrix of the nonlinear sigma model on the symmetric space. Near the point~$\lambda _{1}$ it can be considered as the deformation of the symmetric scattering. In this case it is natural to search the dual theory as sigma model with target space looking as deformed symmetric space. The metric of sigma model on the manifold is subject to very rigid conditions, namely nonlinear renormalization group~(RG) equations \cite{F}. If one has found the solution of RG equations which gives the observables in the sigma model theory, coinciding with that's derived from the factorized $S$-matrix theory one can conclude that the f\/ield theory with the action~$\mathcal{A}(b)$ is dual to the sigma model on the deformed symmetric space. The short distance pattern of such theory can be studied by RG and conformal f\/ield theory (CFT) methods. The agreement of the CFT data, derived from the action $\mathcal{A}(b)$ (considered as a perturbed CFT) with the data derived from RG data for sigma model gives an additional important test for the duality.

The CFT data play an important role in justif\/ication of the third type of the duality. In this case one has the sigma model with singular metric. The nice property of such sigma models is the validity of RG f\/low from the short distances up to the long ones. The RG trajectory relates the non-rational CFT in the ultraviolet (UV) regime with the rational CFT in the infrared (IR). The f\/ield theory dual to sigma model f\/lowing to rational CFT manifests the phenomenon of quantization of the coupling constant.

The large class of two-dimensional quantum f\/ield theories can be considered as perturbed CFTs. In this paper we consider the integrable f\/ield theories which can be formulated as sine-Liouville CFT perturbed by proper f\/ields. In Section \ref{SL-CFT} we describe sine-Liouville CFT and show the duality of this model with Witten's black hole (cigar) CFT. We note that it gives the simple example of the f\/ield/string duality \cite{ADSCFT+++,ADSCFT,ADSCFT+,ADSCFT++}.

In Section \ref{IM-sec} we describe the $\mathbf{W}$-algebra of S-L CFT and show that the enveloping algebra of this $\mathbf{W}$-algebra contains three dif\/ferent Cartan subalgebras. These Cartan subalgebras determine three series of quantum integrals of motion. Each of them is def\/ined by the integrable perturbation of the sine-Liouville CFT. The local integrals of motion are specif\/ied by their densities $P_{s}$ which are local f\/ields with the Lorentz spins $s$ and satisfy the continuity equation $\partial _{\bar{z}}P_{s}=\partial _{z}\Theta _{s-2}$ with a~local f\/ield $\Theta _{s-2}$. Here $\{z,\bar{z}\}$ are the standard light cone (or complex) coordinates.

The factorized scattering for all three types of integrable perturbations theories are constructed in Section~\ref{S-matrices}. The Bethe ansatz (BA) technique is applied to justify that the f\/irst two f\/ield theories have the dual representations, which are available for weak perturbative analysis in dif\/ferent regions of the coupling constant. The phenomenon of fermion-charged boson duality is studied.

To justify the duality of third integrable perturbation with sausage sigma model we study the RG (Ricci f\/low) equations for the metric of this model in Section~\ref{SM}. It is shown that the RG data are in agreement with the data derived by BA method from scattering theory. The test for duality of f\/ield theory with sausage sigma model, based on comparison of RG data at the f\/inite space circle with the thermodynamic Bethe ansatz (TBA) data is done in Section~\ref{Sausage-circle}. In Section~\ref{Metric-singular} we use the same approach for the analysis of sigma model with singular metric. This solution of Ricci f\/low equation describes the RG trajectory from UV to IR regime. The dual f\/ield theory action is conjectured. In Section~\ref{ZN} we describe the massless scattering theory for the RG f\/low described by singular sigma model and TBA equations following from this scattering theory. The TBA equations, RG data and the methods of perturbed CFTs are used to conjecture the duality and to prove the existence of RG f\/low from non-rational CFT in UV regime to rational in IR regime.

The part of the results of this paper presented in Sections~\ref{SL-CFT},~\ref{S-matrices},~\ref{SM} and~\ref{Sausage-circle} were derived and published in collaboration with A.~Zamolodchikov, Al.~Zamolodchikov and E.~Onofri. Some of results presented in Sections~\ref{IM-sec}, \ref{Metric-singular} and~\ref{ZN} are new and where reported in the conference dedicated to the memory of Vadim Knizhnik (IHES, October 2013). The author dedicates this paper to memory of this brilliant scientist.

\section[Sine-Liouville conformal f\/ield theory -- Witten's black hole duality]{Sine-Liouville conformal f\/ield theory --\\ Witten's black hole duality}\label{SL-CFT}

Witten's two-dimensional black hole model \cite{W} is described by the sigma model with action which corresponds to the metric
\begin{gather}
{\rm d}s^{2}=k\big({\rm d}r^{2}+\big(\tanh ^{2}r\big){\rm d}\theta ^{2}\big). \label{bh}
\end{gather}
This model with the dilaton f\/ield $D=\log \big(\cosh ^{2}r\big)$ is described by the CFT with the central charge $c= 2+\frac{6}{k-2}$, which is also known as the coset ${\rm SL}(2,{\mathbb R})/{\rm U}(1)$-parafermionic theory \cite{DVV}. The spectrum of this CFT is
well known and has the form
\begin{gather}
\Delta _{P,m,n}=-\frac{1}{k-2}+P^{2}+\frac{1}{4k}(m\pm nk)^{2}, \label{sp1}
\end{gather}
where $P$ is continuous variable and $m$, $n$ the are integers respectively. The numbers~$m$ and $n$ are called the momentum and the winding numbers. If $\theta$ is $2\pi$ periodic coordinate the metric~(\ref{bh}) describes a manifold with a shape of semi-inf\/inite cigar
\begin{figure}[h!]\centering
\includegraphics[width=.4\textwidth]{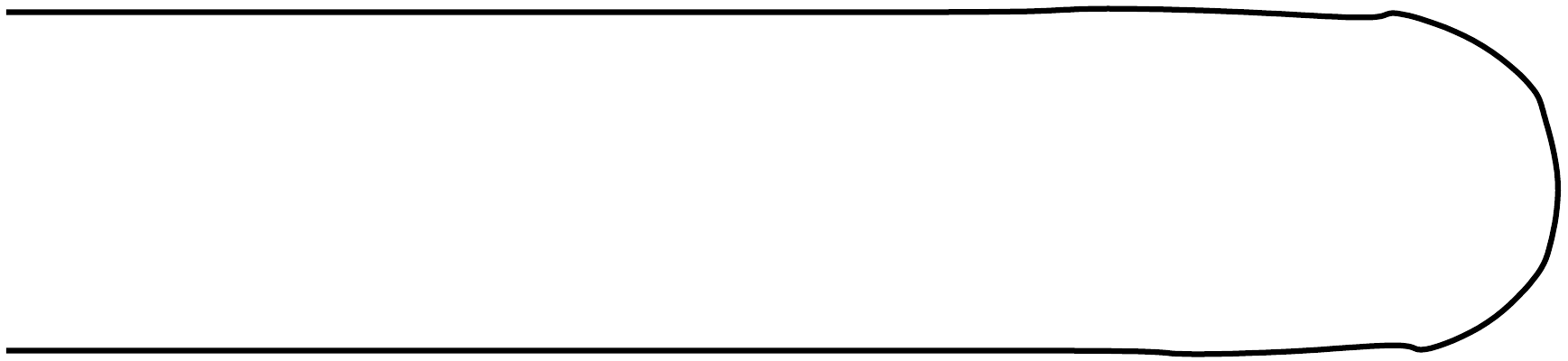}
\caption{Euclidean 2d black hole ``cigar''.}\label{fig1}
\end{figure}

\noindent
One can easily see from this picture that the momentum number $m$ is conserved and the winding number $n$ of the string moving on the cigar can change.

This CFT has a $T$-dual theory with the metric and dilaton, which can be derived from (\ref{bh}) by the transformation $r\rightarrow r+i\frac{\pi}{2}$
\begin{gather*}
{\rm d}s^{2}=k\big({\rm d}r^{2}+\big(\coth ^{2}r\big){\rm d}\hat{\theta}^{2}\big). 
\end{gather*}
The spectrum of the $T$-dual CFT has the same form (\ref{sp1}) with the substitution $m\leftrightarrow n$, i.e., the momentum number transforms to the winding one. Corresponding manifold has a form of the trumpet~\cite{DVV}
\begin{figure}[h!]\centering
\includegraphics[width=.4\textwidth]{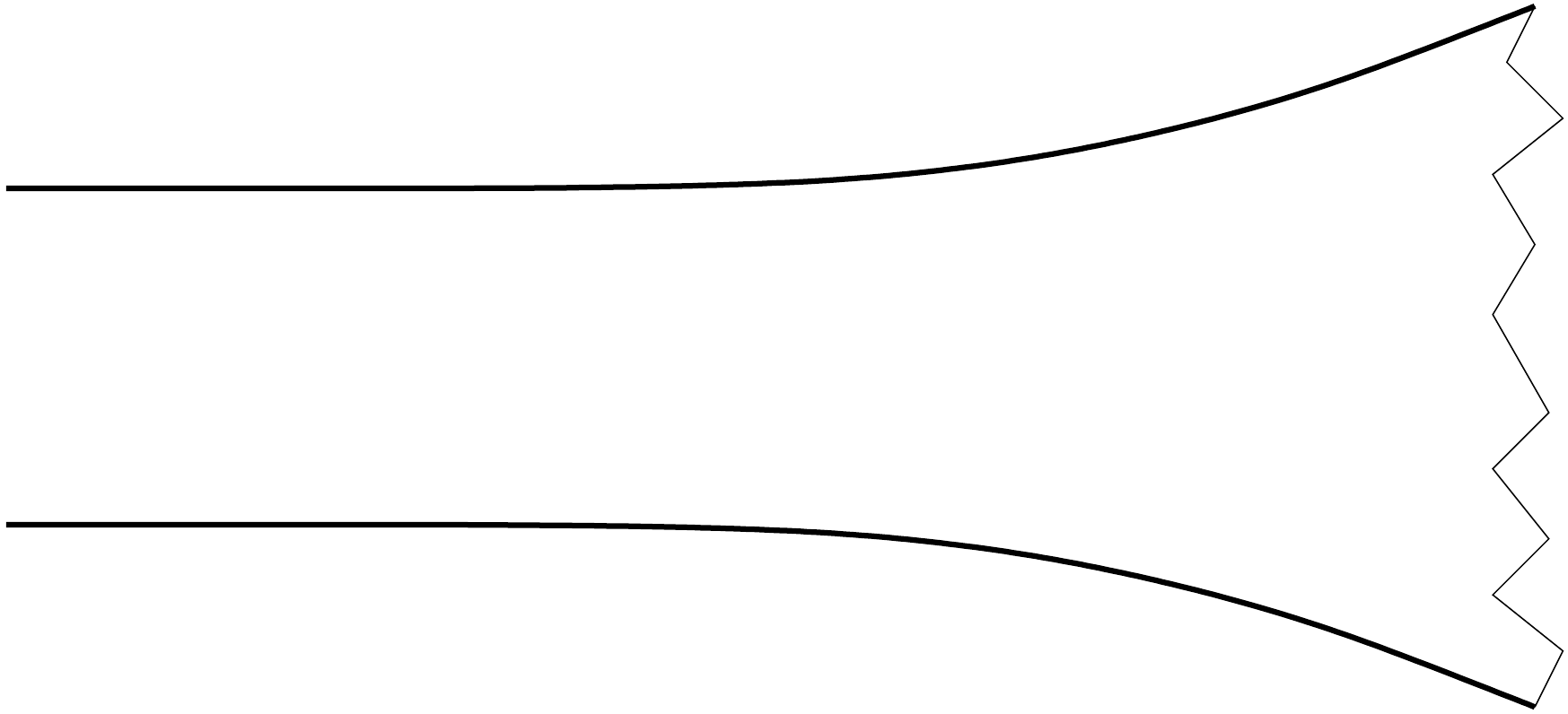}
\caption{``Trumpet'' embedded in 3d Euclidean space.}\label{fig2}
\end{figure}

\noindent
Now one can see that the winding number on this manifold is conserved and the momentum number can change due to the singularity of this manifold.

Sine-Liouville CFT is described by the action
\begin{gather*}
\mathcal{A}=\int {\rm d}^{2}x\left( \frac{(\partial _{\mu}\varphi)^{2}+(\partial _{\mu}\phi)^{2}}{16\pi}+2\mu e^{b\varphi}\cos (a\phi)\right), 
\end{gather*}
where we accept the normalization of the f\/ields $\varphi$, $\phi $
\begin{gather*}
\langle\varphi (z,\bar{z})\varphi (0)\rangle=-2\log ( z\bar{z}) +\cdots,\qquad \langle\phi (z,\bar{z})\phi (0)\rangle=-2\log ( z\bar{z}) +\cdots,
\end{gather*}
and the parameters $a$ and $b$ satisfy the relation
\begin{gather*}
a^{2}-b^{2}=\frac{1}{2}.
\end{gather*}
The stress energy tensor for this CFT is
\begin{gather*}
T=-\frac{1}{4}(\partial _{z}\varphi)^{2}-\frac{1}{4}(\partial _{z}\phi)^{2}+\frac{1}{4b}\partial _{z}^{2}\varphi.
\end{gather*}
If we parametrize: $a^{2}=\frac{k}{4}$, $b^{2}=\frac{k-2}{4}$ the central charge of S-L model will coincide with the central charge of ${\rm SL}(2,{\mathbb R})/{\rm U}(1)$ coset CFT
\begin{gather*}
c=2+\frac{3}{2b^{2}}=2+\frac{6}{k-2}.
\end{gather*}%
The coset ${\rm SL}(2,{\mathbb R})/{\rm U}(1)$ CFT is a parafermionic CFT. The ${\rm SL}(2,{\mathbb R})$ parafermions (non-compact parafermions) $(\Psi,\Psi ^{\ast})\equiv (\Psi ^{(+)},\Psi ^{(-)})$ can be represented~\cite{NEM} by the chiral parts $\phi (z)$ and $\varphi (z)$ of local f\/ields $\phi (z,\bar{z})$, $\varphi (z,\bar{z})$: $\phi (z,\bar{z})=\phi (z)+\bar{\phi}(\bar{z})$,
\begin{gather*}
\Psi ^{(\pm)}(u)=\frac{i}{2a}(ia\partial _{u}\phi \pm b\partial _{u}\varphi)e^{\pm \frac{i}{2a}\phi (u)}.
\end{gather*}
These currents commute with the f\/ields $\mathcal{V}_{\pm}(z)=e^{b\varphi (z)\pm ia\phi (z)}$ which form the potential of sine-Liouville (S-L) CFT, i.e.,
\begin{gather*}
\oint_{u}{\rm d}z\Psi ^{(\pm)}(u)\mathcal{V}_{+}(z)=\oint_{u}{\rm d}z\Psi ^{(\pm)}(u)\mathcal{V}_{-}(z)=0.
\end{gather*}
It means that the $\mathbf{W}$-algebra generated by the holomorphic ${\rm SL}(2,{\mathbb R})/{\rm U}(1)$ parafermionic currents $\Psi ^{(\pm)}(u)$ as the set of local currents appearing in their operator product expansion coincides with the $\mathbf{W}$-algebra of S-L model
\begin{gather}
\Psi ^{(+)}(u)\Psi ^{(-)}(0)=\frac{1}{u^{2+2/k}}\left\{ I+\frac{b^{2}}{a^{2}}u^{2}W_{2}(u)+u^{3}\frac{i}{2a^{3}}W_{3}(u)+\cdots\right\}, \label{ope}
\end{gather}
where $W_{2}=T(z)$ and all other currents $W_{i}$ can be derived from OPE (\ref{ope}). For example
\begin{gather*}
W_{3} =\frac{\big(6b^{2}+1\big)}{6}(\partial \phi)^{3}+b^{2}\partial \phi (\partial \varphi)^{2}+2b^{3}\big(\partial \phi \partial ^{2}\varphi -\partial
\varphi \partial ^{2}\phi \big)- b\partial \varphi \partial ^{2}\phi +a^{2}/3\partial ^{3}\phi. 
\end{gather*}
All currents $W_{i}$ commute with the f\/ields $\mathcal{V}_{\pm}(v)$ and form the symmetry algebra of sine-Liouville CFT. The primary f\/ields of S-L model are the local f\/ields $(\partial _{\mu}\hat{\phi}=\varepsilon _{\mu \nu}\partial _{v}\phi)$
\begin{gather}
\Phi _{\alpha,n,m}=\exp \left( \alpha \varphi +ian\phi +i\frac{1}{4a}m\hat{\phi}\right). \label{prf}
\end{gather}
The right and left dimensions of these f\/ields are $\Delta _{\alpha,n,m}^{(\pm)}=\alpha \big(\frac{1}{2b}-\alpha \big)+\frac{1}{4k}(m\pm nk)^{2}$ and
coincide with the spectrum of the primary f\/ields in ${\rm SL}(2,{\mathbb R})/{\rm U}(1)$ CFT if we put $\alpha -\frac{1}{4b}=iP$. Two point functions $z_{12}^{2\Delta _{\alpha,n,m}^{(+)}}$ $\bar{z}_{12}^{2\Delta _{\alpha,n,m}^{(-)}}\langle\Phi _{\alpha,n,m},\Phi _{\alpha,-n,-m}\rangle$ (ref\/lection amplitudes) of these primary f\/ields in S-L theory $(\alpha ^{\prime}=\alpha -\frac{1}{4b})$ can be easily calculated and are
\begin{gather}
R_{\alpha,n,m} =\left( \frac{\pi \mu}{4b^{2}}\right) ^{-2\alpha ^{\prime
}/b}\frac{\Gamma (1+4b\alpha ^{\prime})\Gamma (1+\alpha ^{\prime}/b)}{\Gamma (1-4b\alpha ^{\prime})\Gamma (1-\alpha ^{\prime}/b)}\nonumber \\
\hphantom{R_{\alpha,n,m} =}{} \times\frac{\Gamma \big(\frac{1}{2}-2b\alpha ^{\prime}+\frac{|m|+nk}{2}\big)\Gamma \big(
\frac{1}{2}-2b\alpha ^{\prime}+\frac{|m|-nk}{2}\big)}{\Gamma \big(\frac{1}{2}+2b\alpha ^{\prime}+\frac{|m|+nk}{2}\big)\Gamma \big(\frac{1}{2}+2b\alpha ^{\prime}+\frac{|m|-nk}{2}\big)},\qquad k=4a^{2}. \label{qra}
\end{gather}
These functions coincide with the two point functions in ${\rm SL}(2,{\mathbb R})/{\rm U}(1)$ CFT. The same is valid for the three point functions. As these functions together with the symmetry algebra $\mathbf{W}$ determine completely the theory, we can conclude (as it was done in the unpublished
paper by A.~Zamolodchikov, Al.~Zamolodchikov and author) that S-L theory is dual to ${\rm SL}(2,{\mathbb R})/{\rm U}(1)$ CFT.

\section[Integrals of motion and integrable perturbations of sine-Liouville CFT]{Integrals of motion and integrable perturbations\\ of sine-Liouville CFT}\label{IM-sec}

The starting point for the analysis of the integrable perturbations of CFT with $\mathbf{W}$ symmetry is the description of dif\/ferent Cartan subalgebras in the enveloping algebra of $\mathbf{W}$. Every such Cartan subalgebra is classif\/ied by integrable perturbation and generates the hierarchy of integrals of motions in the perturbed CFT. These integrals can be represented by their densities $P_{s}$ which are def\/ined up to the total derivative $\partial O$. If we want that our hierarchy started with density of spin three the density should coincide up to derivative with~$W_{3}$. It is convenient to introduce the notation: $T_{1}=-\frac{1}{4}(\partial _{z}\varphi)^{2}-\frac{1}{2}\big(2b+\frac{1}{2b}\big)\partial _{z}^{2}\varphi $ which is formally the stress-energy tensor of Liouville CFT with the coupling constant $-2b$. Then the density $P_{3}^{(1)}$ is
\begin{gather*}
P_{3}^{(1)}=\frac{\big(6b^{2}+1\big)}{6}(\partial \phi)^{3}-4b^{2}\partial \phi T_{1}. 
\end{gather*}
The density of the next integral in this hierarchy has spin $4$ and equals to
\begin{gather*}
P_{4}^{(1)} =\left( \frac{5b^{2}+1}{4}\right) (\partial \phi)^{4}+\left(\frac{8b^{4}+8b^{2}+1}{2}\right) (\partial ^{2}\phi)^{2} +4b^{2}{:}T_{1}^{2}{:}+6b^{2}T_{1}(\partial \phi)^{2}, \label{P41}
\end{gather*}
here ${:}\cdot{:}$ denotes the regular part of the operator product. The densities $P_{s}^{(1)}$ have all integer spins and transform under $\partial \phi \rightarrow -\partial \phi $ as $P_{s}^{(1)}\rightarrow (-)^{s}P_{s}^{(1)}$. The integrals $I_{s}$ of this hierarchy correspond to perturbation of S-L model
by the exponential term $\mu _{1}e^{-2b\varphi}$.

The second hierarchy is generated by the densities $P_{s}^{(2)}$ with even $s $. The f\/irst non-trivial~$P_{4}^{(2)}$ is
\begin{gather*}
P_{4}^{(2)}=P_{4}^{(1)}+9b^{2}\frac{2b^{2}-1}{2b^{2}+3}{:}T^{2}{:}. \label{P42}
\end{gather*}
The densities $P_{s}^{(2)}$ are invariant under the transformation $\partial \phi \rightarrow -\partial \phi $. They correspond to the S-L CFT perturbed by the operator $\mu _{2}e^{-b\varphi}$.

The third hierarchy of densities $P_{s}^{(3)}$ with even spins $s$ is invariant under $\partial \phi \rightarrow -\partial \phi $ and $\partial \varphi \rightarrow -\partial \varphi $. The f\/irst non-trivial $P_{4}^{(3)}$ is
\begin{gather*}
P_{4}^{(3)}=P_{4}^{(1)}+12b^{2}(4b^{2}+1){:}T^{2}{:}. 
\end{gather*}
It corresponds to S-L model perturbed by f\/ields $\mu e^{-b\varphi}\cos a\phi $, and $\mu e^{-b\varphi}\cos a\hat{\phi}$.

\section{Scattering theory and dual representations}\label{S-matrices}

\textbf{1. Integrable perturbation $\boldsymbol{\mu _{1}e^{-2b\varphi}}$ (f\/irst hierarchy).} With the f\/irst integrable perturbation $\mu _{1}e^{-2b\varphi}$ we have
\begin{gather}
L_{1}=\frac{(\partial _{\mu}\varphi)^{2}+(\partial _{\mu}\phi)^{2}}{16\pi}+2\mu e^{b\varphi}\cos (a\phi)+\mu _{1}e^{-2b\varphi}. \label{L1}
\end{gather}
For small $b$ one can use the two-dimensional fermion-boson Coleman--Mandelstam correspondence \cite{C,M}, between f\/ields $\phi $ and $\psi $ to rewrite $L_{1}$ in the form convenient for the perturbation theory (PT) in $b$:
\begin{gather}
L_{1}=\frac{1}{16\pi}\left((\partial _{\mu}\varphi)^{2}+\frac{2M^{2}}{b^{2}}\cosh 2b\varphi +2M\overline{\psi}\psi e^{b\varphi}\right)+L_{\rm TM}, \label{L1F}
\end{gather}
where the term $\frac{M^{2}}{b^{2}}e^{2b\varphi}$ plays the role of the usual counterterm canceling the divergencies coming from the fermion loops and $L_{\rm TM}$ is the Lagrangian of massless Thirring model
\begin{gather}
L_{\rm TM}=\frac{1}{8\pi}\left(i\overline{\psi}\gamma _{\mu}\partial _{\mu}\psi -\frac{b^{2}}{2a^{2}}\big(\overline{\psi}\gamma _{\mu}\psi \big)^{2}\right). \label{TM}
\end{gather}
In \looseness=-1 the PT the spectrum of the theory has two charged particles $\psi$, $\psi ^{+}$ with masses $M$\footnote{The exact relation between the parameter $\mu_{1}$ and the physical mass $M$ in the action~\eqref{L1} can be derived by the BA method \cite{CSH,MMU} and has the form
\begin{gather*}
 \mu_{1}=-\frac{\Gamma(-b^{2})}{\pi\Gamma(1+b^{2})}\left(\frac{M}{4\sqrt{\pi}}\Gamma\left(\frac{1}{2(1+b^{2})}\right)\Gamma\left(1+\frac{b^{2}}{2(1+b^{2})}\right)
 \right)^{2(1+b^{2})}.
\end{gather*}} and one unstable for $b^{2}>0$ bosonic neutral particle with mass $2M$. The integral $P_{3}^{(1)}$ forbids the ref\/lection amplitude $R_{+-}(\theta)$ (here $\theta $ is the relative rapidity of colliding particles) in $\psi \psi ^{+}$ scattering. It means that the scattering is diagonal and is a pure phase. Namely the $S$-matrix~$S^{(1)}$ is
\begin{gather*}
S_{++}(\theta)=S_{--}(\theta)=S_{+-}(i\pi -\theta)=-\frac{\cosh \big(\frac{\theta}{2}+i\frac{\Delta}{2}\big)}{\cosh \big(\frac{\theta}{2}-i\frac{\Delta}{2}\big)}. 
\end{gather*}
The PT gives $\Delta =\pi \frac{b^{2}}{a^{2}}+O\big(b^{6}\big)$. To derive exact relation between $b$ and $\Delta $ we can use BA approach. Our theory has ${\rm U}(1)$ symmetry generated by the charge
\begin{gather}\label{charge}
 Q=\frac{1}{8\pi}\int \psi \psi ^{+}{\rm d}x_{1}.
\end{gather}
We add to our Hamiltonian the term $-AQ$ where $A$ is an external f\/ield (chemical potential) and calculate the asymptotic of the specif\/ic ground state energy $\mathcal{E}(A)$ (GSE) in the limit $\frac{A}{M}\rightarrow \infty$. In this limit we can neglect all terms that contain parameter~$M$ in the Lagrangian and derive the well known expression for the massless Thirring model
\begin{gather*}
\mathcal{E}(A\rightarrow \infty)=-\frac{a^{2}A^{2}}{\pi}=-\frac{\big(1+2b^{2}\big)A^{2}}{2\pi}. 
\end{gather*}
The same value can be calculated from the Bethe ansatz (BA) equations in the external f\/ield. Due to the additional term $AQ$ in the Hamiltonian every positive charged particle acquires the additional energy $A$. For $A>M$ the ground state contains a sea of these particles $\psi (\theta)$ which f\/ill all positive states inside some interval $-B<\theta <B$. The distribution $\epsilon (\theta)$ of these particles is determined by their scattering amplitude $S_{++}(\theta)$. The GSE in the f\/ield $A$ has the form
\begin{gather*}
\mathcal{E}(A)-\mathcal{E}_{0} =-\frac{M}{2\pi}\int \cosh (\theta)\epsilon (\theta){\rm d}\theta, 
\end{gather*}
where $\epsilon (\theta)$ satisf\/ies, inside the interval $-B<\theta <B$, the BA equation
\begin{gather}
\int_{-B}^{B}\hat{K}^{(1)}(\theta -\theta ^{\prime})\epsilon (\theta ^{\prime}){\rm d}\theta ^{\prime}=A-M\cosh \theta \label{ker}
\end{gather}
and $B$ is determined by the boundary conditions $\epsilon (\pm B)=0$. The kernel $\hat{K}^{(1)}(\theta)$ is related with the $\psi \psi$ scattering phase as
\begin{gather}
\hat{K}^{(1)}(\theta)=\delta (\theta)-\frac{1}{2\pi i}\frac{{\rm d}}{{\rm d}\theta} \log ( S_{++}(\theta)). \label{kt}
\end{gather}
It has the Fourier transform
\begin{gather}
K^{(1)}(\omega)=\frac{2\sinh [ (\pi -\Delta)\omega /2] \cosh[ (\pi +\Delta)\omega /2]}{\sinh ( \pi \omega)}.\label{ko}
\end{gather}
The asymptotic $\mathcal{E}(A\rightarrow \infty)$ can be expressed through the kernel $K(\omega)$ at $\omega=0$ \cite{FOZ}. For the kernel~\eqref{ko} one has
\begin{gather*}
\mathcal{E}(A\rightarrow \infty)=-\frac{A^{2}}{2\pi K(0)}=-\frac{A^{2}}{2(\pi -\Delta)},
\end{gather*}%
or $\Delta =\frac{\pi b^{2}}{a^{2}}=\frac{2\pi b^{2}}{1+2b^{2}}$. At $b\rightarrow \infty $, $\Delta \rightarrow \pi $. We introduce parameter $%
4\gamma ^{2}=\frac{1}{b^{2}}$. Then $\Delta =\pi -\frac{2\gamma ^{2}\pi}{1+2\gamma ^{2}}$ and
\begin{gather*}
S_{++}(\theta)=S_{--}(\theta)=S_{+-}(i\pi -\theta)=\frac{\sinh \big(\frac{\theta}{2}-\frac{i\gamma ^{2}\pi}{1+2\gamma ^{2}}\big)}{\sinh \big( \frac{\theta}{2}+\frac{i\gamma ^{2}\pi}{1+2\gamma ^{2}}\big)}.
\end{gather*}
In the limit $b\rightarrow \infty$, $\gamma \rightarrow 0$ we have the free theory, but contrary to $b\ll 1$ case $S$-matrix tends to $1$ but not to~$-1$. Such behavior is characteristic for bosonic particles. In this limit $K^{(1)}(\omega)\rightarrow \frac{(\pi -\Delta)\omega}{\tanh \pi \omega}$. The BA equations with this kernel can be solved and
\begin{gather}
\mathcal{E}(A)-\mathcal{E}(0) =-\frac{(A-M)^{2}}{2(\pi -\Delta)}+O(1). \label{eb}
\end{gather}
The GSE (\ref{eb}) has a threshold behavior ($A\rightarrow M$) unusual for fermionic particles, which have there the singularity $(A-M)^{3/2}$. The quadratic behavior \eqref{eb} is characteristic behavior for weakly coupled bosonic particles, it ref\/lects the instability of free bosons under the introduction of external f\/ield. The $S$-matrix in this limit coincides with $S$-matrix of complex sinh-Gordon model with
\begin{gather}
\hat{L}_{1}=\frac{1}{4\pi}\left( \frac{\partial _{\mu}\chi \partial _{\mu}\chi ^{\ast}}{1+\gamma ^{2}\chi \chi ^{\ast}}+M^{2}\chi \chi ^{\ast}\right), \label{CSG}
\end{gather}
where $\chi =\chi _{1}+ i \chi _{2}$ is a complex scalar f\/ield. Complex sinh-Gordon model is integrable classically~\cite{LR, PL}. The quantum
integrability and renormalizability of this theory was studied in~\cite{DD, HEC}. For $b\gg 1$, $\gamma \ll 1$ the theory is described by weakly coupled charged bosons $\chi$, $\chi ^{\ast}$ with masses~$M$. The external f\/ield $A$ can be introduced by $\partial _{0}\rightarrow \partial _{0}+iA$. It is easy to check that~$\mathcal{E}(A)$ in this case can be derived by minimization of Euclidean action for constant~$\chi $
\begin{gather*}
\mathcal{E}(A)=\min_{\chi}\left[ \frac{1}{4\pi}\left( -\frac{A^{2}\chi \chi ^{\ast}}{1+\gamma ^{2}\chi \chi ^{\ast}}+M^{2}\chi \chi
^{\ast}\right) \right]. 
\end{gather*}

The duality between the QFTs (\ref{L1F}) and (\ref{CSG}) gives us an example of the fermion-boson duality, when the charged particles being the fermions in weak coupling regime become the bosons in the strong coupling one.

\textbf{2. Integrable perturbation $\boldsymbol{\mu _{2}e^{-b\varphi}}$ (second hierarchy).} The Lagrangian $L_{2}$ in the form convenient for PT at small $b$ is
\begin{gather}
L_{2}=\frac{1}{16\pi}(\partial _{\mu}\varphi)^{2}+2M\overline{\psi}\psi e^{b\varphi}+\frac{M^{2}}{b^{2}}\big(e^{2b\varphi}+2e^{-b\varphi}\big)+L_{\rm TM}, \label{L2F}
\end{gather}
where $L_{\rm TM}$ is the Lagrangian for the massless Thirring model~\eqref{TM}. In the PT the spectrum consists from two charged particles $\psi$, $\psi ^{+}$ with masses~$M$ and one neutral particle (their bound state) with mass~$\sqrt{3}M$. In this case we do not have conserved currents with odd spins and in PT amplitude $R_{+-}\neq 0$. The ${\rm U}(1)$ symmetric solution of Yang--Baxter equation up to CDD factors coincides with the $S$-matrix of the sine-Gordon model~$S_{\rm SG}$:
\begin{gather}
S_{++}^{++}(\theta)=S_{+-}^{+-}(i\pi -\theta)=-e^{i\delta _{\lambda}(\theta)},\qquad S_{-+}^{+-}(\theta)=\frac{-i\sin \pi \lambda
}{\sinh \lambda (i\pi -\theta)}e^{i\delta _{\lambda}(\theta)},\label{SG}
\end{gather}
where
\begin{gather*}
\delta _{\lambda}=\int_{0}^{\infty}{\rm d}\omega \frac{\sin (\omega \theta)\sinh [ \omega \pi (1-\lambda)/2\lambda]}{\omega \cosh
(\omega \pi /2)\sinh [ \omega \pi (1/2\lambda)]}, 
\end{gather*}
$\lambda =\frac{1}{\beta _{\rm SG}^{2}}-1$, and $\beta _{\rm SG}$ is the coupling constant in sine-Gordon model. $S$-matrix (\ref{SG}) corresponds to $c_{\rm UV}=1$. We expect that for our QFT $c_{\rm UV}=2$. It can be achieved by addition of one CDD factor. At $b\rightarrow 0$ the $S$-matrix for QFT \eqref{L2F} is $-I+O (b^{2})$. It means that at~$b\rightarrow 0$ the CDD factor should cancel sine-Gordon $S$-matrix, i.e., at~$b=0$ $S$-matrix $S_{\rm SG}$ should be diagonal at~$\lambda (0)$ and contain only one factor. It happens if $\lambda (0)=3$. For this $\lambda$ the $S$-matrix~(\ref{SG}) has the poles at $\theta _{1}=\frac{i2\pi}{3}$ and $\theta _{1}=\frac{i\pi}{3}$. The f\/irst pole will be canceled by CDD factor and the second gives the bound state with the mass $M_{1}=2M\sin \frac{\pi}{3}=\sqrt{3}M$, what agrees with the PT. It means that
\begin{gather*}
S^{(2)}(\theta)=-\frac{\sinh (\theta)-i\sin (\pi /\lambda)}{\sinh (\theta)+i\sin (\pi /\lambda)}S_{\rm SG}(\lambda,\theta). 
\end{gather*}
To f\/ind the function $\lambda (b)$ we introduce the f\/ield $A$ coupled with the charge~\eqref{charge}. At $A/M\rightarrow \infty $ we have the same result that for QFT (\ref{L1F}) $\mathcal{E}(A)\rightarrow -\frac{a^{2}A^{2}}{\pi}$. The kernel $\hat{K}^{(2)} (\theta) $ is now expressed throw $\frac{1}{2\pi}\frac{{\rm d}}{{\rm d}\theta}\delta ^{++}(\theta)$, where $\delta ^{++}(\theta)$ is the phase of scattering of particles $\psi \psi \colon \delta ^{++}(\theta)=\delta _{\lambda}+\frac{1}{i}\log \frac{\sinh (\theta)-i\sin (\pi /\lambda)}{\sinh (\theta)+i\sin (\pi /\lambda)}$. The Fourier transform of this kernel is
\begin{gather*}
K^{(2)}(\omega)=\frac{\sinh \big(\frac{\pi \omega}{2}\big(\frac{\lambda -1}{\lambda}\big)\big)\cosh \big(\frac{3\pi \omega}{2\lambda}\big)}{\sinh \big(\frac{\pi \omega}{\lambda}\big)\cosh \big(\frac{\pi \omega}{2}\big)} 
\end{gather*}
and
\begin{gather*}
\mathcal{E}(A\rightarrow \infty)=-\frac{A^{2}}{2\pi K^{(2)}(0)}=-\frac{A^{2}}{\pi (\lambda -1)}.
\end{gather*}
Comparing two expressions we derive $\lambda =\frac{3+2b^{2}}{1+2b^{2}}$. At $b\rightarrow \infty$, $\lambda \rightarrow 1$ and we again have $S\rightarrow I+O(\gamma ^{2})$. The mass of the neutral particle $M_{1}=2M\sin \big(\frac{\pi}{\lambda}\big) $ and at $\lambda =2$ or $2b^{2}=1$ it disappears from the spectrum and we have only two charged particles. At $b^{2}\gg 1,\gamma ^{2}=\frac{1}{4b^{2}}$ the kernel $K^{(2)}=2\gamma ^{2}\frac{\pi \omega \cosh (3\pi \omega /2)}{\cosh (\pi \omega /2)\sinh (\pi \omega)}$ and $\mathcal{E}(A)$ can be calculated. The result can be written in the parametric form~\cite{INDT}
\begin{gather*}
\mathcal{E}(A)-\mathcal{E}(0)=-\frac{A^{2}}{4\pi \gamma ^{2}}\left( 1-3X+X^{2}+\frac{X^{3}}{(1-X)}\right), \qquad
\left(\frac{M}{A}\right) ^{2} =\frac{4X^{3}}{(1-X)}.
\end{gather*}
It is easy to check that $\mathcal{E}(A)$ coincides with the minimum of the Euclidean action (after $\partial _{0}\rightarrow \partial _{0}+iA $) for the QFT with Lagrangian
\begin{gather*}
\hat{L}_{2}=\frac{1}{4\pi}\left( \frac{\partial _{\mu}\chi \partial _{\mu}\chi ^{\ast}}{1+\gamma ^{2}\chi \chi ^{\ast}}+M^{2}\chi \chi ^{\ast}\big(1+\gamma ^{2}\chi \chi ^{\ast}\big)\right).
\end{gather*}
The perturbative expansion in $\gamma ^{2}$ for the $S$-matrix conf\/irms this suggestion. Here we see not only the phenomenon of the Dirac fermion-charged boson duality but also that due to the non-trivial kinetic term, the interaction, which looks as repulsive potential becomes attractive for f\/inite $\gamma $ and produce the bound state for $\gamma ^{2}>\frac{1}{2}$.

\textbf{3. Integrable perturbation $\boldsymbol{\mu e^{-b\varphi}\cos (a\phi)}$ dual to the sausage model (third hierarchy).} The Lagrangian of the perturbed S-L CFT now is
\begin{gather}
L_{3}=\frac{1}{16\pi}\big((\partial _{\mu}\varphi)^{2}+(\partial _{\mu}\phi)^{2}\big)+4\mu \cosh (b\varphi)\cos (a\phi). \label{L3}
\end{gather}
The Lagrangian $L_{3}$ in the form convenient for PT at small $b$ is
\begin{gather}
L_{3}=\frac{1}{16\pi}\left((\partial _{\mu}\varphi)^{2}+2M\overline{\psi}\psi \cosh (b\varphi)+\frac{M^{2}}{b^{2}}\sinh ^{2}(b\varphi)\right)+L_{\rm TM}, \label{L3F}
\end{gather}
where the term with $\sinh ^{2}b\varphi $ plays the role of the counterterm and $L_{\rm TM}$ is given by~\eqref{TM}. In the PT the spectrum consists from two charged $(\psi,\psi ^{+})$ or $( +,-)$ and one neutral particle $(\varphi)$ or $(0) $ with the same mass~$M$. The $S$-matrix $S^{(3)}(\theta)$ for such set of particles is known~\cite{FOZ, FZ} and up to $\mathbf{C}$, $\textbf{P}$, $\textbf{T}$
\begin{gather*}
S_{kl}^{ij}(\theta)=S_{\bar{k}\bar{l}}^{\bar{\imath}\bar{j}}(\theta)=S_{lk}^{ji}(\theta)=S_{ij}^{kl}(\theta)
\end{gather*}
and crossing
\begin{gather*}
S_{kl}^{ij}(\theta)=S_{l\bar{\imath}}^{j\bar{k}}(i\pi -\theta)
\end{gather*}
symmetries has the following independent amplitudes $S_{kl}^{ij}(\theta)$, where $i,j,k,l=+,0,-$, and $i+j=k+l$, $\bar{j}=-j$,
\begin{gather}
S_{++}^{++}(\theta) = \frac{\sinh \lambda (\theta -i\pi)}{\sinh \lambda (\theta +i\pi)},\qquad
S_{+0}^{0+}=\frac{-iS_{++}^{++}(\theta)\sin 2\pi \lambda}{\sinh \lambda (\theta -2i\pi)},\nonumber\\
S_{+0}^{+0}=\frac{S_{++}^{++}(\theta)\sinh\lambda \theta}{\sinh \lambda (\theta -2i\pi)}, \qquad
S_{-+}^{+-} =\frac{-\sin 2\pi \lambda \sin \pi \lambda}{\sinh \lambda (\theta -2i\pi)},\qquad S_{00}^{00}=S_{+0}^{+0}+S_{-+}^{+-}.\label{FST}
\end{gather}
In the perturbation theory we derive $\lambda =\frac{1}{2}-b^{2}+O(b^{4})$. To f\/ind exact function $\lambda (b)$ we introduce the external f\/ield $A$ coupled with the charge~\eqref{charge}. Again we have that $\mathcal{E}(A\rightarrow \infty)\rightarrow -\frac{a^{2}A^{2}}{\pi}$. The kernel of BA equations (\ref{ker}), (\ref{kt}) is now
\begin{gather*}
\hat{K}^{(3)}(\theta)=\delta (\theta)-\frac{1}{2\pi i}\frac{{\rm d}}{{\rm d}\theta}\log ( S_{++}^{++}(\theta)) =\delta (\theta)-\frac{\lambda}{\pi}\frac{\sin 2\pi \lambda}{\cosh 2\lambda \theta -\cos 2\pi \lambda}.
\end{gather*}
It has the Fourier transform
\begin{gather}
K^{(3)}(\omega)=\frac{2\sinh \big[ \frac{\pi \omega}{2}\big] \cosh \big[ \frac{\pi \omega (1-\lambda)}{2\lambda}\big]}{\sinh \big[ \frac{\pi \omega}{2\lambda}\big]}, \label{k3}
\end{gather}
and
\begin{gather*}
\mathcal{E}(A\rightarrow \infty)=-\frac{A^{2}}{2\pi K^{(3)}(0)}=-\frac{A^{2}}{4\pi \lambda},
\end{gather*}%
or $\lambda =\frac{1}{4a^{2}}=\frac{1}{2(1+2b^{2})}=\frac{1}{k}$. At $b\rightarrow \infty$, $\lambda \rightarrow 0$ and our $S$-matrix coincides with the $S$-matrix of the ${\rm O}(3)$ sigma model. For f\/inite $b\gg 1$ it is natural to search the dual representation of the QFT (\ref{L3}), (\ref{L3F}) as ${\rm U}(1)$ symmetric deformation of this sigma model. As the deformed ${\rm O}(3)$ sigma model should be compared with QFT~(\ref{L3}),~(\ref{L3F}), we consider the BA equations here more attentively. We should show that observables in this QFT calculated from the $S$-matrix data in the UV region coincide with the same observables in the sigma-model analysis. As the observables calculated from $S$-matrix data we consider here the GSE in the external f\/ield $\mathcal{E}(A)$ and $E_{0}(R)$ the GSE of the model at the f\/inite size circle of length $2\pi R$ (see later). To start, we consider the function $\mathcal{E}(A)$. The BA equations can be solved by generalized Winner--Hopf technique \cite{GNN}, which permits to develop the large $\big( \frac{A}{M}\big) $ expansion. This expansion of $\mathcal{E}(A)$ for kernel (\ref{k3}) runs in two types of exponents: instanton exponents $\big(\frac{M}{A}\big)^{2q}$ and perturbative exponents $\big(\frac{M}{A}\big)^{\frac{2\lambda}{1-\lambda}}$, namely
\begin{gather}
\mathcal{E}(A)=-\frac{A^{2}}{4\pi \lambda}\sum_{q=0}^{\infty}\left( \frac{M}{A}\right) ^{2q}f^{(q)}\left( \frac{A}{M}\right), \label{exp}
\end{gather}
where the functions $f^{(q)}\big( \frac{A}{M}\big)$, are the regular series in $\big(\frac{M}{A}\big)^{\frac{2\lambda}{1-\lambda}}$. For example,
\begin{gather*}
f^{(0)}\left( \frac{A}{M}\right) =\left( 1-4\left( \frac{1-\lambda}{1-2\lambda}\right) ^{2}\frac{\Gamma \big( \frac{-\lambda}{2-2\lambda} \big) \Gamma \big( \frac{1}{2-2\lambda}\big)}{\Gamma \big( \frac{\lambda}{2-2\lambda}\big) \Gamma \big( \frac{-1}{2-2\lambda}\big)}\left( \frac{2\lambda M}{A}\right) ^{\frac{2\lambda}{1-\lambda}}+\cdots \right).
\end{gather*}
The instanton exponents in the expansion (\ref{exp}) appear due to the instanton contributions and the perturbative exponents as the sum of PT around the $q$-instanton solution. The instanton exponents do not depend on coupling constant~$b$. We note that instantons appear in all sigma models with the compact two-dimensional target space.

For $b\gg 1$, $\lambda \ll 1$ and the main contribution to $\mathcal{E}(A)$ comes from $f^{(0)}\big( \frac{A}{M}\big) $. The BA equations simplify drastically in the scaling limit $\lambda \rightarrow 0$, $\log \big(\frac{A}{M}\big) \rightarrow \infty $ with $\lambda \log \big( \frac{A}{M}\big)$ f\/ixed. Corrections to the scaling behavior also can be developed. Here we give the scaling limit of $\mathcal{E}(A)$ together with the leading
(``two-loop'') correction~\cite{FOZ}
\begin{gather}
\mathcal{E}(A)=-\frac{A^{2}}{4\pi \lambda}\frac{1-\mathrm{q}}{1+\mathrm{q}}\left( 1+4\lambda \frac{\mathrm{q}}{1-\mathrm{q}^{2}}\log \left( \frac{1-\mathrm{q}}{2\lambda}\right) +O\big(\lambda ^{2}\log ^{2}\lambda \big)\right),\label{es}
\end{gather}
where $\mathrm{q}=\big( \frac{Me^{3/2}}{8A}\big)^{\frac{2\lambda}{1-\lambda}}$. In the limit $\lambda \rightarrow 0$ we recover the result of~\cite{HMN} for ${\rm O}(3)$~SM.

\section{Sigma models, Ricci f\/low and sausage sigma model}\label{SM}
The nonlinear sigma models (SM) in two-dimensional space-time are widely used in QFT as well as in relation with string theory. They are described by the action
\begin{gather}
\mathcal{A}[G] =\frac{1}{4\pi}\int G_{ij}(X)\partial _{\mu}X^{i}\partial _{\mu}X^{j}{\rm d}^{2}x+\cdots, \label{metr}
\end{gather}%
where $X^{i}$ are coordinates in $d$-dimensional manifold called target space and symmetric mat\-rix~$G_{ij}(X)$ is the corresponding metric.

The standard approach to two-dimensional SM is the perturbation theory. If the curvature is small one can use the following renormalization group (RG)-evolution equation \cite{F}. Let $t$ be the RG time (the logarithm of the scale) $t\rightarrow -\infty $ in UV limit and $t\rightarrow \infty $ in IR. Then the one loop RG evolution equation is
\begin{gather}
\frac{{\rm d}}{{\rm d}t}G_{ij}=-R_{ij}+O\big(R^{2}\big), \label{Ricci}
\end{gather}%
where $R_{ij}$ is the Ricci tensor of $G$.

The analysis of this equation shows \cite{GP} that in general the nonlinear evolution equation is unstable in the sense that even if one starts from manifold of small curvature everywhere at some scale~$t^{\ast}$, under evolution in both directions $t\rightarrow \pm \infty $ the metric~$G(t) $ develops at least some regions where its curvature grows and~(\ref{Ricci}) is no more applicable. If it happens in the UV direction $t\rightarrow -\infty $ the action~(\ref{metr}) does not def\/ines any local QFT. However, special solutions exist where UV direction is stable and curvature remains small up to $t\rightarrow -\infty $, permitting one to def\/ine the local QFT (at least perturbatively). For example, if we have homogeneous symmetric space, its metric grows in the UV and curvature monotonously decreases and we are dealing with an UV asymptotically free QFT unambiguously def\/ined by the action~(\ref{metr}). Very interesting class of the solutions of Ricci f\/low equation form the solutions related with deformed symmetric spaces. The simple example of such solution (deformed sphere~$S^{2}$ or sausage) is considered later. The asymptotic of the solutions of Ricci f\/low equations at $t\rightarrow -\infty $ correspond to the f\/ixed points of Ricci f\/low. They are more symmetric and subject to methods of CFT.

To study the large distance physics one should f\/ind a suitable approach. The quantum integrability is one of the most successful lines in studying non-critical SMs. The quantum integrability and global symmetries of the metric are manifested in the factorized scattering theory (FST) of corresponding excitations. The FST is rather rigid and its internal restriction does not permit a wide variety of consistent constructions. The FST contains all the information about background integrable QFT. The methods of integrable QFTs allows one to compute some of\/f-mass-shell observables on the base of FST. In the UV region these observables should be compared with that's following from SM~(\ref{metr}). If they match non-trivially in the UV region it is naturally to suggest the chosen FST as the scattering theory of integrable SM. Moreover, one can use the FST as a non-perturbative def\/inition of the SM.

At $d=2$ the Ricci f\/low equation (\ref{Ricci}) is much simplif\/ied. There one has $R_{ij}=\frac{1}{2}\mathcal{R}G_{ij}$, where $\mathcal{R}$ is the scalar curvature. Then one can always choose the conformal coordinates such that $G_{ij}=e^{\Phi}\delta _{ij}$ and $\mathcal{R}=-e^{-\Phi}\partial_{X^{i}}\partial _{X^{i}}\Phi $. The equation~(\ref{Ricci}) now reads
\begin{gather*}
-\frac{{\rm d}}{{\rm d}t}\Phi =\frac{1}{2}\mathcal{R}+\cdots \qquad \text{or}\qquad \frac{{\rm d}}{{\rm d}t}\Phi =e^{-\Phi}\partial _{X^{i}}\partial _{X^{i}}\Phi +\cdots, 
\end{gather*}
the two loop correction to the f\/irst of these equations is $\frac{1}{4}\mathcal{R}^{2}$ \cite{F}.

Our SM is ${\rm U}(1)$ or axially symmetric, so in conformal coordinates $X$, $Y$ we choose $0\leq X<2\pi $ as angular coordinate and $\Phi (Y)$ independent on~$X$. Then one loop equation looks as nonlinear heat equation
\begin{gather*}
\frac{{\rm d}}{{\rm d}t}\Phi (Y) =e^{-\Phi}\partial _{Y}^{2}\Phi (Y).
\end{gather*}
The solution of this equation is \cite{FOZ}
\begin{gather*}
e^{\Phi}=\frac{\sinh 2\nu (t_{0}-t)}{\nu (\cosh 2\nu (t_{0}-t)+\cosh 2Y)},
\end{gather*}
where $\nu $ is the real parameter. It corresponds to the action
\begin{gather}
\mathcal{A}_{\rm SSM}=\frac{1}{4\pi}\int \frac{((\partial _{\mu}X)^{2}+(\partial _{\mu}Y)^{2})\sinh 2\nu (t_{0}-t)}{\nu (\cosh 2\nu (t_{0}-t)+\cosh 2Y)}{\rm d}^{2}x. \label{sa}
\end{gather}
This manifold for $u=\nu (t_{0}-t)\ll 1$ looks like a sphere and at $u\gg 1$ we see a long sausage of length $L\simeq \sqrt{2\nu}(t_{0}-t)$, and in the middle it tends to the cylinder with radius $\sqrt{1/\nu}$. For this reason we call this QFT as sausage sigma model (SSM). For $u=\nu (t_{0}-t)\gg 1$ the sausage looks as two long cigars~(\ref{bh}) glued together. In particular, it means that in the UV region $u\gg 1$ one can use the black hole (with $k=\frac{1}{\nu}$) or S-L CFT data for analysis of SSM.

\begin{figure}[t]\centering
\includegraphics[width=.5\textwidth]{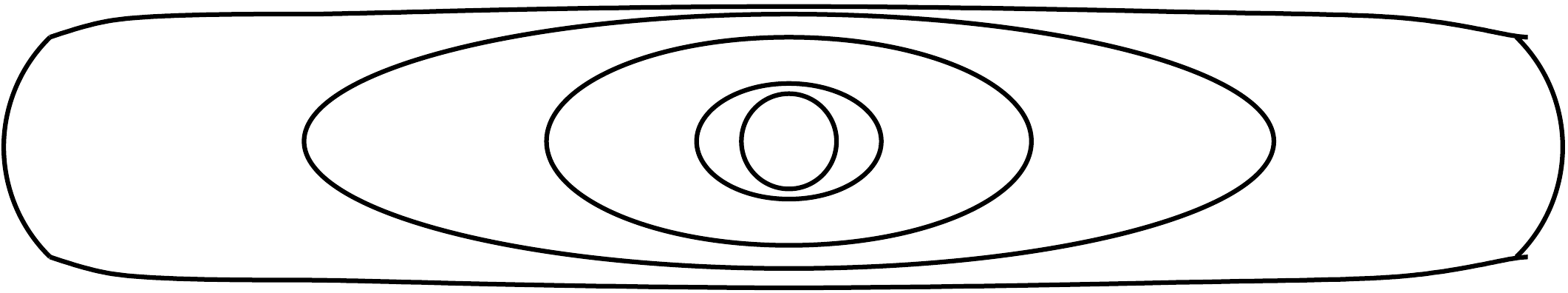}
\caption{RG evolution of sausage.}\label{fig3}
\end{figure}

The SSM possesses the instantons and one can add to the action~(\ref{sa}) topological term $i\theta _{T}T$ where $\theta _{T}$ is a topological angle. The instantons play important role at large distances and for $\theta_{T}=\pi $ this theory f\/low from the CFT with $c_{\rm UV}=2$ to the CFT with $c_{\rm IR}=1$. The full analysis of this theory was done in~\cite{FOZ}. Here we will be interested in the UV behavior, which does not depend on~$\theta _{T}$.

One can compare the GSE in the external f\/ield $A$ derived from FST data and the same value derived from SSM~(\ref{sa}). The f\/ield $A$ introduces the scale $(t_{0}-t)=\log \frac{A}{M}$. The introduction of the f\/ield $A$ amounts the substitution $\partial _{0}X\rightarrow (\partial _{0}+iA)X$ in the action~(\ref{sa}). The GSE derived from the action of SSM corresponds to the minimum of Euclidean action. This minimum is achieved at $Y=0$. The one loop GSE $\mathcal{E}(A)$ then is
\begin{gather}
\mathcal{E}(A)=-\frac{1}{4\pi \nu}A^{2}(\tanh (u)+O(\nu \log \nu)),\qquad u=\nu \log \left( \frac{A}{M}\right). \label{esc}
\end{gather}
Comparing GSEs (\ref{es}) and (\ref{esc}) we derive that they coincide in the scaling (one-loop) appro\-xi\-mation and that $\nu =\lambda =\frac{1}{k}$.

It is worth mentioning that the SSM action (\ref{sa}) admits also a simple parametrization in terms of unit-vector $n_{j}(x)$ on $S^{2}$ in which one can easily see SSM as the deformation of ${\rm O}(3)$-sigma model
\begin{gather}
\mathcal{A}_{\rm SSM}=\frac{1}{4\pi g(t)} \int\sum_{j=1}^{3}\frac{(\partial _{\mu}n_{j}) ^{2}}{1-\frac{\nu ^{2}}{2g^{2}(t)}n_{3}^{2}}{\rm d}^{2}x,\label{o3}
\end{gather}
where $g(t)=\nu \cot (\nu (t_{0}-t))$.

\section{Sausage at the circle}\label{Sausage-circle}
In this section we consider SM at the circle of length $R$. It introduces the scale $(t_{0}-t)=\log \big( \frac{1}{RM}\big)$. In the UV scaling regime: $-\log (RM)\rightarrow \infty$, $\nu \rightarrow 0$ such that $u=-\nu \log (RM)$ is f\/inite, our one-loop approximation is exact up to $O(\nu \log \nu)$. In this approximation we can use minisuperspace approach to calculate the UV corrections to GSE~$E_{0}(R)$, ef\/fective central charge $E_{0}(R)=-\frac{\pi}{6R}c(R)$ and to the energies $E_{i}(R)$ of exited states. It was in shown in \cite{FO, FOZ} that these values can be expressed throw the eigenvalues of the covariant operator $\hat{h}$
\begin{gather}
\hat{h}=-\nabla _{t}^{2}+\frac{1}{4}\mathcal{R}_{t},\qquad \hat{h}\Psi _{i}=\frac{e_{i}(R)}{6}\Psi _{i}, \label{mss}
\end{gather}
where $\nabla _{t}^{2}$ is the Laplace operator and $\mathcal{R}_{t}$ is the scalar curvature in the SSM metric renormalized at the scale $R$. Then with the accuracy $\nu \log \nu $
\begin{gather}
c(R)=2-e_{0}(R),\qquad E_{i}(R)=E_{0}(R)+\frac{\pi (e_{i}-e_{0})}{R}.\label{m1}
\end{gather}
Operator $\hat{h}$ is self-adjoint with respect to the scalar product with the SSM metric
\begin{gather*}
( \Psi _{1},\Psi _{2}) =\int \Psi _{1}^{\ast}\Psi _{2}e^{\Phi(y)}{\rm d}x{\rm d}y,
\end{gather*}
where coordinates $x$, $y$ can be considered as the zero modes of the f\/ields $X$, $Y$.

It is easy to see that the operator $\hat{h}/\nu $
\begin{gather*}
\frac{1}{\nu}\hat{h}\Psi =-\frac{e^{\Phi (y)}}{\nu}\left(\frac{1}{2}\frac{{\rm d}^{2}}{{\rm d}y^{2}}+\frac{1}{8}\Phi ^{\prime \prime}(y)\right)\Psi
\end{gather*}
depends only on the scaling variable $u=-\nu \log (RM)$. It means that the eigenvalues $e_{i}(R)$ scale as $\nu e_{i}(u)$. We can search for the solution $\Psi =e^{ixm}\Psi _{m}$. After the substitution
\begin{gather*}
e^{y-u}=\frac{\operatorname{cn}(z|s)}{\operatorname{sn}(z|s)},\qquad \psi _{m}=\sqrt{\frac{\operatorname{sn}(z|s)\operatorname{cn}(z|s)}{\operatorname{dn}(z|s)}}\Psi _{m}
\end{gather*}
with modulus of the elliptic Jacobi function $s^{2}=1-e^{-4u}$ the equation can be written in the Lam\'{e} form
\begin{gather}
\left( -\frac{{\rm d}^{2}}{{\rm d}z^{2}}-\frac{\operatorname{cn}^{2}(2z|s)}{\operatorname{sn}^{2}(2z|s)} +\frac{m^{2}\operatorname{dn}^{2}(z|s)}{\operatorname{sn}^{2}(z|s)\operatorname{cn}^{2}(z|s)}\right) \psi _{m}=\frac{\kappa_{m,j}s^{2}}{6}\psi _{m}, \label{lame}
\end{gather}
where $e_{m,j}(R)=\nu \kappa _{m,j}(u)$ and the boundary conditions for the solutions are $\psi _{m}\sim z^{m+\frac{1}{2}}$ at $z\rightarrow 0$, $\psi _{m}\sim (K-z)^{m+\frac{1}{2}}$ at $z\rightarrow K$, where $K(s^{2}) $ is a real period of Jacobi functions.

For small $u$, $s^{2}\simeq 4u$, $K\simeq \frac{\pi}{2}$, the equation (\ref{lame}) can be easily solved
\begin{gather*}
\psi _{m}=(\sin 2z)^{m+1/2}P_{j}(\cos 2z),
\end{gather*}
where $P_{j}$ are Legendre polynomials. We derive $\frac{\kappa _{m,j}}{6}\simeq \frac{j(j+1) +1/2}{u}$, $j\geq m$ and $e_{0}(R)=\nu \kappa
_{m,0}=\frac{3}{\log (1/RM)}$. This asymptotic is universal for all spheres $S^{d}$ with $d>1$
\begin{gather*}
c(R)=d-\frac{3}{2}d/\log (1/RM)+O\big( \log ( \log (1/RM)) /\log ^{2}(1/RM)\big).
\end{gather*}
We consider now another limit $u\gg 1$, $s^{2}\rightarrow 1$, $K\simeq 2u+\log 4$. In this limit the potential $V(z)$ in the Lam\'{e} equation with
exponential accuracy looks as
\begin{gather*}
V_{l}(z) =-\frac{1}{\sinh ^{2}2z}+m^{2}\coth ^{2}z,\qquad 0<z\ll K,\\ 
V_{r} ( z_{1}) =-\frac{1}{\sinh ^{2}2z_{1}}+m^{2}\coth ^{2}z_{1},\qquad 0<z_{1}=K-z\ll K.
\end{gather*}
We parametrize $\frac{\kappa _{m,n}}{6}=m^{2}+4p^{2}$. Then in the middle one can neglect the potential term and~$\psi _{m}$ is the plane wave solution. At the left and right ends $z\sim 0$, $z\sim K$ the equation can be solved exactly in terms of the hypergeometric functions $F( A,B,C,z)$
\begin{gather}
\psi _{m}^{(l)} =N_{l}(p,m) (\tanh z) ^{m+\frac{1}{2}}(\cosh z) ^{2ip}F\big( A,A,m+1,\tanh ^{2}z\big), \nonumber\\
\psi _{m}^{(r)} =\psi _{m}^{(l)}(z_{1}),\qquad\text{where}\quad A=\frac{m+1-2ip}{2}. \label{ps}
\end{gather}
The constant $N_{l}(p,m)=N_{r}(p,m)$ is chosen from the condition
\begin{gather*}
\psi _{m}^{(l)}\simeq e^{2ipz}+\mathrm{R}_{l}^{(cl)}(p,m) e^{-2ipz},\qquad z\gg 1,\qquad \psi _{m}^{(r)}\simeq e^{2ipz_{1}}+\mathrm{R}_{r}^{(cl)}(p,m) e^{-2ipz_{1}}.
\end{gather*}%
The corresponding solutions (\ref{ps}) are specif\/ied by the ref\/lection amplitudes
\begin{gather}
\mathrm{R}_{l}^{(cl)}=\mathrm{R}_{r}^{(cl)}=\frac{\Gamma (1+4ip)\Gamma ^{2}\big(\frac{1}{2}-2ip+\frac{|m|}{2}\big)}{\Gamma (1-4ip)\Gamma ^{2}\big(\frac{1}{2}+2ip+\frac{|m|}{2}\big)}.\label{rlr}
\end{gather}
These amplitudes coincide with the semiclassical limit $b\gg 1$, $b\alpha ^{\prime}=ip$, of CFT ref\/lection amplitudes~(\ref{qra}) with $n=0$. (For $n\neq 0 $ the energy levels are very large and our minisuperspace approach does not work.) Matching the solutions in dif\/ferent domains we derive
\begin{gather}
\frac{1}{6}\kappa _{m,j}=m^{2}+\frac{\pi ^{2}(j+1)^{2}}{4(u+r_{m})^{2}}+O\big(u^{-5}\big),\nonumber\\ r_{m}=\psi (1)-\psi \left(\frac{m+1}{2}\right),\qquad \psi(x) =\frac{\Gamma ^{\prime}(x)}{\Gamma (x)}. \label{spec}
\end{gather}
The UV asymptotics with the accuracy $O\big(M^{2}R^{2}\log MR\big)$ can be derived from exact CFT ref\/lection amplitudes with $n=0$. The potential terms in $L_{3}$ are $2\mu e^{b\varphi}\cos a\phi $, and $2\mu e^{-b\varphi}\cos a\phi $. Both of them correspond to S-L CFT and have the same
``quantum'' ref\/lection amplitudes (\ref{qra}). To write the equation for UV asymptotics of $e_{m,n}(R)$ \cite{ALZZ} we
should make the substitution $\mu \rightarrow \mu (\frac{R}{2\pi})^{2-2\Delta _{\rm Pot}}$ and take the exact relation between $\mu $ and $M$.
In our case $\Delta_{\rm Pot}=a^{2}-b^{2}=\frac{1}{2}$ and exact relation between $\mu $ and $M$ can be derived by BA method and is $\mu =\frac{M}{2\pi}$. The equation for the levels is $\mathrm{R}_{l}^{(q)}\mathrm{R}_{r}^{(q)}=1$. Namely, $e_{m,j}(R)=6\big(\frac{m^{2}}{k}+4P_{m,j}^{2}\big)$, where $P_{m,j}$ are the solutions to the equation: $\log \mathrm{R}_{l}^{(q)}\mathrm{R}_{r}^{(q)}=2i\pi (j+1)$ with $\mathrm{R}_{l}^{(q)}=\mathrm{R}_{r}^{(q)}=\mathrm{R}^{(q)}$,
\begin{gather*}
\mathrm{R}^{(q)}=\left( \frac{MR}{16\pi b^{2}}\right) ^{-2iP/b} \frac{\Gamma (1+4biP)\Gamma (1+iP/b)\Gamma ^{2}\big(\frac{1}{2}-2biP+\frac{|m|}{2}\big)}{\Gamma (1-4biP)\Gamma (1-iP/b)\Gamma ^{2}\big(\frac{1}{2}+2biP+\frac{|m|}{2}\big)}. 
\end{gather*}
It is easy to check that in the scaling limit the UV asymptotics coincide with that derived by minisuperspace approach. The two loop correction to~(\ref{spec}) can be easily calculated for $u\gg 1$ using $\mathrm{R}^{(q)}$. It is
\begin{gather*}
\frac{\nu e_{m,j}}{6}=m^{2}+\frac{\pi ^{2}(j+1)^{2}}{4(u+r_{m})^{2}}\left( 1+\frac{\nu}{2}\frac{\log \big( \frac{\nu}{4\pi}\big) +\psi (1)}{(u+r_{m})}+O\big( \nu ^{2}\log ^{2}\nu \big) \right).
\end{gather*}

The ef\/fective central charge can be calculated with arbitrary accuracy from $S$-matrix data (TBA equations). The TBA equations for $\lambda =\nu =\frac{1}{k}=\frac{1}{N}$ form the system of $N+1$ coupled nonlinear equations for $N+1$ functions $\varepsilon _{a}(\beta)$\footnote{For arbitrary $\lambda<\frac{1}{2}$ the calculation of the GSE and the energies of excited states can be also derived from the nonlinear integral equations (see \cite{Ahn:2017mff,Bazhanov:2017nzh} for details).}:
\begin{gather}
R\rho _{a}(\beta)=\varepsilon _{a}(\beta)+\frac{1}{2\pi}\int \sum_{b=0}^{N}\frac{l_{ab}}{\cosh (\beta -\beta ^{\prime})}\log \big[
1+e^{-\varepsilon _{b}(\beta ^{\prime})}\big] {\rm d}\beta ^{\prime},\label{tbe}\\
E_{0}(R)=-\frac{1}{2\pi}\int \sum_{b=0}^{N}\rho _{b}(\beta)\log \big[1+e^{-\varepsilon _{b}(\beta)}\big]{\rm d}\beta, \nonumber 
\end{gather}
where $l_{ab}$ is the incidence matrix of the af\/f\/ine $D_{N}$ Dynkin diagram and $\rho _{a}=M\delta _{0}^{a}\cosh \beta $ for SSM at $\ \theta _{T}=0$ and $\rho _{a}=\frac{M\delta _{0}^{a}}{2}e^{\beta}+\frac{M\delta _{1}^{a}}{2}e^{-\beta}$ for SSM with topological term at $\theta _{T}=\pi$.
\begin{figure}[t]\centering
 \includegraphics{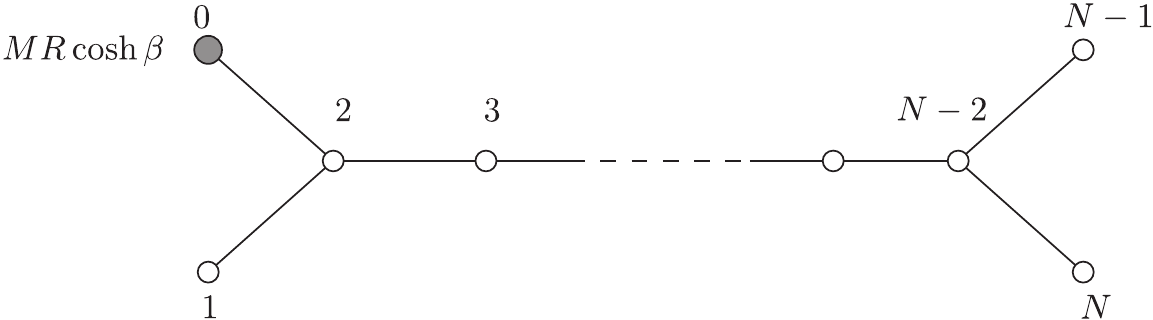}
\caption{Incidence diagram of TBA systems and source term for SSM.}\label{fig4}
\end{figure}
The TBA calculations reproduce with great accuracy the function $e_{0}(R)$ and scaling function $\kappa _{m,n}(u)$. The comparison of numerically computed from equation~(\ref{lame}) function $\kappa _{0}(u)$ and function $\frac{1}{N}e_{0}(R)$ derived from TBA equations is represented in~\cite{FOZ}. The excellent agreement of UV behavior of observables derived from FST data~(\ref{FST}) for QFT~(\ref{L3F}) with that's derived from the Ricci f\/low data for SSM (\ref{sa}) give us a reason to conjecture that these theories coincide and are dual.

\section[Sigma model with singular metric and RG f\/low to rational CFT]{Sigma model with singular metric\\ and RG f\/low to rational CFT}\label{Metric-singular}

In the previous sections we discussed the SMs with compact target space. One can easily derive from the Ricci f\/low equation (\ref{Ricci}) that at $d=2$ the volume of this manifold
\begin{gather}
\Omega =\int \sqrt{G}{\rm d}^{2}X=-2(t-t_{0}) \label{v}
\end{gather}
grows linearly for $t\rightarrow -\infty $. Contrary the ``forward'' RG evolution always ends at some point~$t_{0}$ where manifold shrinks to a point and curvature becomes inf\/inite. It means that the only possibility to have non-trivial RG evolution in the range $-\infty <(t-t_{0}) <\infty $ is to work with metric where integral (\ref{v}) does no exist. It happens for the non-compact manifolds with singular metric. Here we consider this interesting possibility.

The action of the sausage model (\ref{sa}) admits the analytic continuation $Y\rightarrow Y+\frac{i\pi}{4}$, $u\rightarrow u+\frac{i\pi}{4}$,
\begin{gather}
\mathcal{A}_{\rm MSM}=\frac{1}{4\pi}\int \frac{((\partial _{\mu}X)^{2}+(\partial _{\mu}Y)^{2})\cosh 2u}{\nu (\sinh 2u+\sinh 2Y)}{\rm d}^{2}x. \label{am}
\end{gather}
This metric has singularity at $Y=-u$, i.e., coordinate $Y$ in target space should be considered in the region $Y>-u$. The metric is singular but all geodesic distances are f\/inite and we can apply to the analysis of this ``massless'' sigma model (MSM) the minisuperspace approach. The part of corresponding manifold which can be embedded to Euclidean space looks as a bell for large negative $u$ (IR regime) and as a surface surgery of cigar with trumpet for $u\gg 1$ (UV regime).
\begin{figure}[t]\centering
 \includegraphics[width=.9\textwidth]{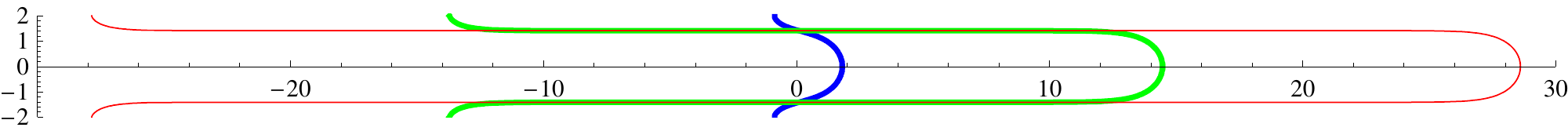}
\caption{Ricci f\/low from the bell in IR to hybrid of cigar and trumpet in UV.}\label{fig5}
\end{figure}

The metric of the bell is
\begin{gather*}
{\rm d}s^{2}=k\big({\rm d}r^{2}+\big(\tan ^{2}r\big){\rm d}\theta ^{2}\big),
\end{gather*}
where $k=\frac{1}{\nu}$. It was shown in \cite{K} that the action with this metric for integer $k=N$ descri\-bes~$Z_{N}$ parafermionic CFT~\cite{FZAM}. The analysis of the CFT with this action was done in~\cite{MMS}. There was shown that the theory is consistent only for integer $k=N$ and the ${\rm U}(1)$ symmetry of action is broken up to group~$Z_{N}$. It means that in quantum case the surgery is possible only for $\nu =\frac{1}{N}$.

The MSS equations with metric (\ref{am}) have the form (\ref{mss}), (\ref{m1}). After the substitution $\Psi =e^{ixm}\Psi _{m}$, $e^{y-u}=\frac{\operatorname{dn} ( z|s)}{k\operatorname{sn}(z|s)}$, $\psi _{m}=\sqrt{\frac{\operatorname{sn}(z|s)\operatorname{dn}(z|s)}{\operatorname{cn}(z|s)}}\Psi _{m}$, where $s^{2}=\frac{1}{1+e^{-4u}}$, it has again the Lam\'{e} form
\begin{gather}
\left( -\frac{{\rm d}^{2}}{{\rm d}z^{2}}-\frac{\operatorname{dn}^{2}(2z|s)}{\operatorname{sn}^{2}(2z|s)} +\frac{m^{2}\operatorname{cn}^{2}(z|s)}{\operatorname{sn}^{2}(z|s)\operatorname{dn}^{2}(z|s)}\right) \psi _{m,j}=\frac{\kappa
_{m,j}^{\prime}}{6}\psi _{m,j}, \label{l1}
\end{gather}%
here $e_{m,j}(R)=\nu \kappa _{m,j}^{\prime}(u)$. This transformation maps the point $y=\infty $ to $z=0$ and $y=-u$ to $z=K(s^{2})$, and $\psi
_{m}\sim z^{m+\frac{1}{2}}$ at $z\rightarrow 0$, $\psi _{m}\sim (K-z)^{\frac{1}{2}}$ at $z\rightarrow K$.

In the IR limit $u\rightarrow -\infty $, $K\rightarrow \frac{\pi}{2}$, this equation can be solved exactly in terms of the Jacobi polynomials $(\cos z)^{m}\sin 2zP_{j}^{(m,0)}(\cos 2z)$. When the IR limit of RG is described by CFT, the values $\Delta _{i}=\frac{\left( e_{i}-e_{0}\right)}{24}$ coincide with the spectrum of conformal dimensions of primary f\/ields and $d-e_{0}$ with the central charge of CFT. In our case
\begin{gather*}
\Delta _{j,m}=\frac{\nu \big( \kappa _{m,j}^{\prime}-\kappa _{0,0}^{\prime}\big)}{24}=\frac{j(j+1)}{N}-\frac{m^{2}}{4N},\qquad j\leq m,\qquad e_{0}=\frac{6}{N}. 
\end{gather*}

These values correspond in one loop approximation to the spectrum of $Z_{N}$-parafermionic CFT. It can be proved \cite{FO} that perturbation theory in $s^{2}=\frac{1}{1+e^{-4u}}$ for eigenvalues converges for all real~$u$. The eigenvalues $\kappa _{m,j}^{\prime}$ can be expanded in the series in parameter $e^{4u}=(MR)^{-4/N}$. We will see later that corresponding quantum (all loops) series have the IR expansion parameter $(MR)^{-\frac{4}{N+2}}$. For example,
\begin{gather}
\frac{\kappa _{m,j}^{\prime}-\kappa _{0,0}^{\prime}}{6} =4j(j+1) -m^{2}-\frac{(4j(j+1) -m^{2})^{2}}{8j(j+1)}s^{2}+\cdots, \label{dcl} \\
\kappa _{0,0}^{\prime} =6-s^{4}-\frac{1}{2}s^{6}+\cdots =6-e^{8u}+\frac{3}{2}e^{12u}+\cdots. \label{d0}
\end{gather}

In the opposite UV limit $u\gg 1$ the potential term in (\ref{l1}) with exponential accuracy has the form
\begin{gather}
V_{l}(z) =-\frac{1}{\sinh ^{2}2z}+m^{2}\coth ^{2}z,\qquad 0<z\ll K, \nonumber\\
V_{r}( z_{1}) =-\frac{1}{\sinh ^{2}2z_{1}}+m^{2}\tanh^{2}z_{1},\qquad 0<z_{1}=K-z\ll K. \label{pt}
\end{gather}
In the right region the potential $V$ is attractive for $m>0$ and has a~bound states
\begin{gather}
\psi_{m,j}=(\tanh z_{1})^{\frac{1}{2}}(\cosh z_{1})^{j-m+1}F\big({-}j,-j+m,m-2j,\cosh^{-2}z_{1}\big),\\
\frac{1}{6}\kappa _{m,j}^{\prime}=m^{2}-(2j+1-m)^{2},\qquad j=0,\ldots\leq \frac{m-1}{2}.\label{ds}
\end{gather}
These states describe discrete degrees of freedom of the manifold which survive in the UV limit. We note that UV limit is described by the ${\rm SL}(2,{\mathbb R})/{\rm U}(1)$ CFT which was studied in \cite{DVV}. The levels~(\ref{ds}) correspond to discrete series representations of ${\rm SL}(2,{\mathbb R})$ and play an essential role for string theory interpretation of the coset CFT.

In the left region the potential $V$ is repulsive and for $j>\frac{m-1}{2}$ the spectrum can be derived by matching the exact solutions at the left and right ends
\begin{gather*}
\psi _{m}^{(l)} =N_{l}(\tanh z) ^{m+\frac{1}{2}}( \cosh z) ^{2ip}F\big( A,A,m+1,\tanh ^{2}z\big), \\
\psi _{m}^{(r)} =N_{r} ( \tanh z_{1}) ^{\frac{1}{2}} ( \cosh z_{1}) ^{2ip}F\big( A,A-m,1,\tanh ^{2}z_{1}\big) 
\end{gather*}
(here as before $A=\frac{m+1-2ip}{2}$) with the plane wave in the middle. The ref\/lection amplitude $\mathrm{R}_{l}^{(cl)}$ will be again (\ref{rlr})\ and $\mathrm{R}_{r}^{(cl)}$\ is now
\begin{gather}
\mathrm{R}_{r}^{\prime (cl)}=\frac{\Gamma (1+4ip)\Gamma \big(\frac{1}{2}-2ip+\frac{m}{2}\big)\Gamma \big(\frac{1}{2}-2ip-\frac{m}{2}\big)}{\Gamma (1-4ip)\Gamma \big(\frac{1}{2}+2ip+\frac{m}{2}\big)\Gamma \big(\frac{1}{2}+2ip-\frac{m}{2}\big)}. \label{RS}
\end{gather}
The matching leads to
\begin{gather}
\frac{\kappa _{m,j}^{\prime}(u)}{6}=m^{2}+\frac{\pi ^{2}(2j-m+2)}{16(u+r_{m})^{2}}+O\big(1/u^{5}\big). \label{cs}
\end{gather}

\begin{figure}[t]\centering
\includegraphics[width=.57\textwidth]{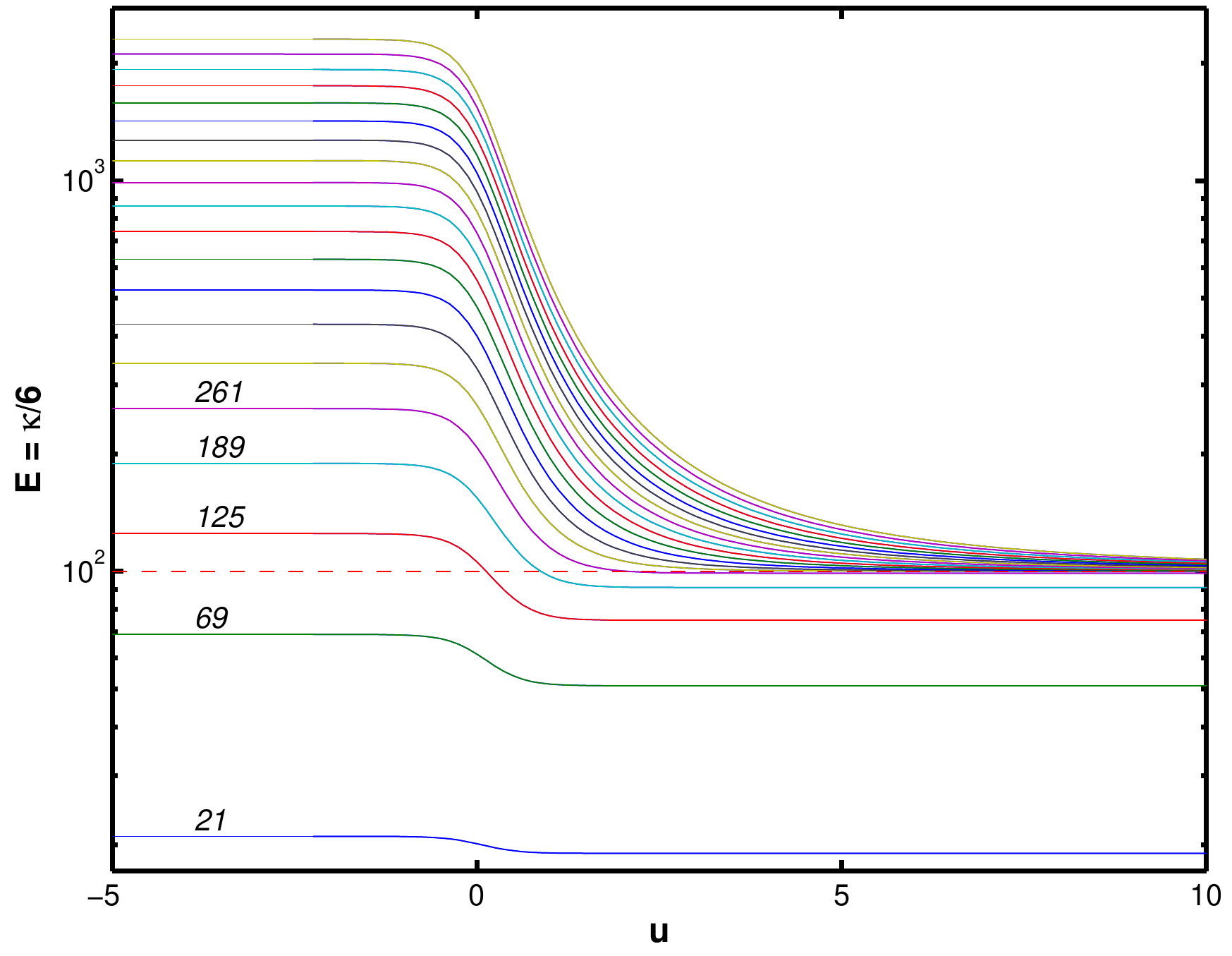}
\caption{Ricci f\/low of the levels from IR to UV for $m=10$.}\label{fig6}
\end{figure}

The f\/low of the spectrum for $m=10$ from IR to UV, i.e., from discrete spectrum for $u\rightarrow -\infty $ to discrete (\ref{ds}) and continuum (\ref{cs}) for $u\rightarrow -\infty $ is shown on the Fig.~\ref{fig6}. One can see that not only the ground state level $e_{0}(R)$ (related with ef\/fective central charge $c(R)=2-e_{0}(R)$) is according to Zamolodchikov's $c$-theorem the decreasing (non-increasing) function of $u$, but all levels also possess this property.

The quantum version of the ref\/lection amplitude (\ref{RS}) is
\begin{gather*}
\mathrm{R}_{r}^{(q)} =\left( \frac{MR}{16\pi b^{2}}\right)^{-2iP/b}\frac{\Gamma (1+4biP)\Gamma (1+iP/b)}{\Gamma (1-4biP)\Gamma (1-iP/b)}\frac{\Gamma \big(\frac{1}{2}-2biP+\frac{m}{2}\big)\Gamma \big(\frac{1}{2}-2biP-\frac{m}{2}\big)}{\Gamma \big(\frac{1}{2}-2biP+\frac{m}{2}\big)\Gamma \big(\frac{1}{2}-2biP-\frac{m}{2}\big)}. 
\end{gather*}
The poles of $\mathrm{R}_{r}^{(q)}(P,m)$ at the physical strip $iP\geq 0$ at the points $P_{j,m}=\frac{i}{4b}(m-2j-1)$, $j\leq \frac{m-1}{2}$, determine exact bound states levels $\big(2b^{2}=2a^{2}-1=\frac{1}{2\nu}-1\big)${\samepage
\begin{gather*}
\frac{1}{6}e_{m,j}=\left(\frac{m^{2}}{k}+4P_{m,j}^{2}\right)=\nu \left( m^{2}-\frac{1}{1-2\nu}(m-2j-1)\right)
\end{gather*}
in agreement with (\ref{ds}) up to $O(\nu ^{2})$.}

The minisuperspace approach is valid for the quantum numbers $j,m\ll k=4a^{2}$. In the UV limit MSM looks as a cigar matched with a trumpet. The CFTs corresponding to these sigma models are T dual. Both of these models are dual to S-L CFT, but interpretation of the primary f\/ields (\ref{prf}) is dif\/ferent. The numbers $m$, $n$ (momentum~$m$ and winding quantum number $n$ for cigar) transform to $n$, $m$ (momentum~$n$ and winding quantum number $m$ for trumpet). This transformation corresponds to $\phi \rightarrow \hat{\phi}$, where $\partial _{\mu}\hat{\phi}=\varepsilon _{\mu \nu}\partial _{v}\phi $. The ref\/lection amplitudes $\mathrm{R}_{l}^{(q)}(p,m)$ and $\mathrm{R}_{r}^{(q)}(p,m) $ correspond to the momentum number
equal to $m$ and winding number equal to $0$ for cigar and vice versa for trumpet. It means that left ($\varphi \rightarrow \infty $) and right ($\varphi \rightarrow -\infty $) CFTs are related by transformation $\phi \rightarrow \hat{\phi}$, i.e., the potential term in the action of QFT dual to MSM is
\begin{gather*}
2\mu e^{b\varphi}\cos (a\phi)+2\mu e^{-b\varphi}\cos \big(a\hat{\phi}\big).
\end{gather*}
But these two terms are mutually local only if $4 a^{2}=k$ is integer, i.e., $k=N$, and we derive again the quantization of the coupling constant.

\section[$Z_{N}$-parafermionic CFT and IR action for Ricci f\/low]{$\boldsymbol{Z_{N}}$-parafermionic CFT and IR action for Ricci f\/low}\label{ZN}
$Z_{N}$-parafermionic CFT \cite{FZAM} with the central charge $c_{N}=2-\frac{6}{N+2}$ describes the critical behavior of $Z_{N}$-Ising model~\cite{ZN}. It has the primary f\/ields $\phi _{j,m}$, $j\leq m$, with the conformal dimensions
\begin{gather*}
\Delta _{j,m}^{\rm (CFT)}=\frac{j(j+1)}{N+2}-\frac{m^{2}}{4N}. 
\end{gather*}
Besides the parafermionic symmetry it has also the symmetry generated by $W_{N}$ algebra. The $Z_{N}$-neutral f\/ields are thermal operators $\phi _{j,0}$ with $\Delta^{\rm (CFT)}_{j,0}= \frac{j(j+1)}{N+2}$, $j=1,\ldots \leq \lbrack N/2]$. To show that the Ricci f\/low, studied in the previous section, in the IR regime is described by the perturbation of the $Z_{N}$-parafermionic CFT, one should f\/ind the receiving operator in the space of the $Z_{N}$-neutral f\/ields of this theory. The perturbative calculations with this IR relevant f\/ield must be consistent with RG results.

Besides the parafermionic symmetry $Z_{N}$-parafermionic CFT has the symmetry algebra~$W_{N}$. This symmetry algebra is generated by the holomorphic f\/ield~$W_{s}(z)$ which appear in the OPE of the parafermionic currents $\psi (z)$ and $\psi ^{\ast}(z)$ (see~\cite{VFN}).

The f\/irst thermal operator $\varepsilon =$ $\phi _{j,0}$ has the descendent f\/ield $E=\mathcal{N}W_{-1}\overline{W}_{-1}\varepsilon $ with dimension $1+\frac{2}{N+2}$. The constant $\mathcal{N}$ provides the conformal normalization of the f\/ield $E$. This f\/ield is unique self dual (with respect $Z_{N}\otimes \hat{Z}_{N}$ or order-disorder duality) integrable IR relevant perturbation in $Z_{N}$-parafermionic CFT. It means that the action describing our f\/low should be written as
\begin{gather}
\mathcal{A}_{\rm IR}=\mathcal{A}_{\rm PF}+\varkappa \int E{\rm d}^{2}x, \label{IRA}
\end{gather}
where $\mathcal{A}_{\rm PF}$ is the action of $Z_{N}$-parafermionic CFT and $\varkappa $ is the coupling constant. The exact relation between $\varkappa $ and mass scale $M$, which appear later in FST and TBA equations, can be derived by BA method~\cite{MMUF, MMU} and is
\begin{gather}
(\pi \varkappa)^{2}=\frac{N^{2}(N-2)^{2}}{3(N+2)^{2}(N+4)^{2}}G(w)G(3w)\left[ \frac{4}{wM}\right] ^{8w}, \label{EX}
\end{gather}%
where $G(w)=\frac{\Gamma (1+w)}{\Gamma (1-w)}$ and $w=\frac{1}{N+2}$.

One can calculate the IR corrections to the spectrum $\Delta _{j,m}^{\rm (CFT)}$ and to the ef\/fective central charge $c(R) $ with the action (\ref{IRA}),~(\ref{EX}). The f\/irst exact corrections to $D_{j,m}=(N+2)\Delta _{j,m}^{\rm (CFT)}$ is
\begin{gather}
D_{j,m}(R)=D_{j,m}-\frac{1}{4}D_{j,m}^{2}\frac{B(j,w)}{j(j+1)}\left( \frac{8\pi}{wMR}\right) ^{4w}+\cdots, \label{DJM}
\end{gather}
where
\begin{gather*}
B( j,w) =\frac{2N^{2}G^{2}(w) G( (2j+2) w)}{(N+2)^{2}G(2w) G( 2jw)}.
\end{gather*}

The exact corrections to the central charge calculated with the action $\mathcal{A}_{\rm IR}$ are
\begin{gather}
c=2-\frac{6}{N+2}+C_{1}(w)\left( \frac{8\pi}{wMR}\right) ^{8v}+C_{2}(w)\left( \frac{8\pi}{wMR}\right) ^{12v}+\cdots, \label{cR}
\end{gather}
where
\begin{gather*}
C_{1}(w)=\frac{wN^{2}(N-2)^{2}G(w) G(3w)}{(N+2)^{2}(N+4)^{2}G(4w) G(-2w)}, \\ C_{2}(w)=\frac{-w3N^{4}(N-4)^{2}G(2w) G(4w)}{2(N+4)^{4}(N+8)^{2}G (6w) G(-3w)}.
\end{gather*}

At $N\gg 1$, $w=\frac{1}{N+2}\ll 1$ the coef\/f\/icients $B_{j,m}$, $C_{1}$, $C_{2}$ are $B_{j,m}=2+O(w)$, $C_{1}(w)=1+O(w)$, $C_{3}(w)=-\frac{3}{2}+O(w) $. Comparing equations~(\ref{DJM}), (\ref{cR}) with (\ref{dcl}), (\ref{d0}) we see their coincidence in one loop approximation.

To make more non-trivial test one can use the TBA equations for MSM. These equations can be derived from the massless FST, consistent with QFTs~(\ref{am}),~(\ref{IRA}). Massless FSTs are widely used for the analysis of RG f\/lows from one critical point in UV to another critical point in IR (see, for example, \cite{FOZ, AAZZ}). The massless particles or kinks in such FST are of two kinds: right-moving with the dispersion low: $\mathrm{e}=\mathrm{p}= \frac{M}{2}e^{\theta}$ and left-moving: $\mathrm{e}=-\mathrm{p}=\frac{M}{2}e^{-\theta}$. The full scattering matrix contains amplitudes $S_{\rm RR}(\theta)$, $S_{\rm LL}(\theta)$ and $S_{\rm RL}(\theta)$ of right-right, left-left and right-left moving particles.

$Z_{N}$-parafermionic CFT which is the basic object for IR action (\ref{IRA}) possesses the symmetry with quantum group ${\rm SL}(2) _{q}$ with $q=\exp \big( \frac{2\pi i}{N+2}\big)$. As the parameter of symmetry~$q$ takes only discrete values it is natural to expect that the scattering matrix will correspond to the massless kinks. The most suitable factorized scattering matrix of kinks with ${\rm SL}(2) _{q}$ symmetry is described in~\cite{VFN}, where all amplitudes of kinks scattering are written explicitly. In~\cite{VFN} this scattering matrix was used for massive kinks in some integrable model also related with $Z_{N}$-parafermionic CFT. Here one should use the same amplitudes for the scattering of right-right, left-left and right-left moving kinks.

The TBA equations can be derived by the procedure of diagonalization of the transfer-matrix with the elements depending on the scattering amplitudes. The massless FST described above leads to the same equations~(\ref{tbe}) for $N+1$ functions $\varepsilon _{a}$, but the source term $\rho _{a}$ is dif\/ferent. Now it is $\rho _{a}=\frac{M\delta _{0}^{a}}{2}e^{\beta}+\frac{M\delta _{N}^{a}}{2}e^{-\beta}$.
\begin{figure}[t]\centering
 \includegraphics{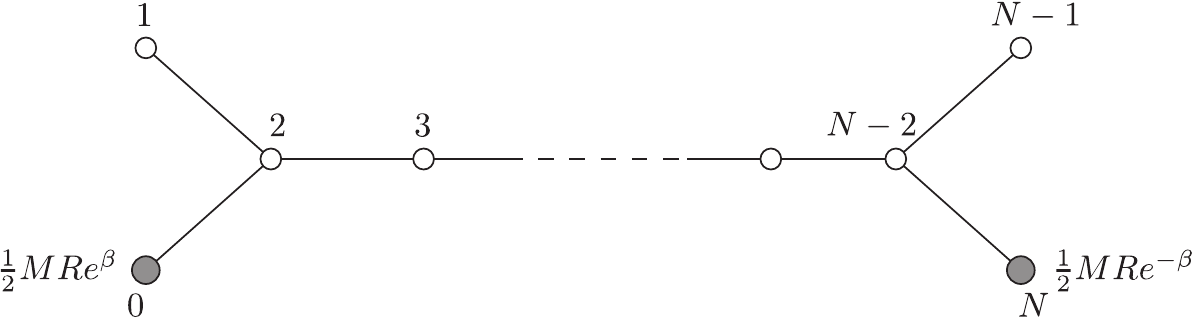}
\caption{Incidence diagram of TBA system and source terms for MSM.}\label{fig7}
\end{figure}

The numerical solution of TBA equations perfectly reproduces the UV and the IR behavior of MSM in the scaling limit (see \cite{FO}). The exact coef\/f\/icients $B( j,w)$, $C_{1}(w)$, $C_{2}(w)$ derived from the action (\ref{IRA}) coincide with all possible numerical accuracy with the corresponding coef\/f\/icients calculated from the TBA equations.

\section{Concluding remarks}\label{Conl}
\begin{enumerate}\itemsep=-1.5pt
\item Sine-Liouville CFT possesses integrable perturbations, corresponding to massive and massless QFTs. All these perturbations establish non-trivial duality properties, like Dirac fermion-charged boson duality, duality with sigma models, which describe ``massive'' and ``massless'' Ricci f\/lows.

\item Sine-Liouville CFT can be generalized to sine-Toda CFT. This CFT possesses three dif\/ferent integrable perturbations~\cite{INDT}. These integrable f\/ield theories have the Lagrangian description in terms of massive Thirring model coupled with non-simply laced af\/f\/ine Toda theories of rank~$r$. Perturbative calculations, analysis of FST and BA technique show that these QFTs have the dual description available for the perturbative analysis in the strong coupling regime. The dual QFTs are formulated in terms of completely bosonic theories, namely, complex sinh-Gordon theory coupled with Toda theories corresponding to dual af\/f\/ine algebras but with smaller rank $\hat{r}=r-1$.

\item The SSM can be generalized to sigma-model with 3d target space~\cite{TWP}. The metric of this sigma model is the deformation of~${\rm O}(4)$ (or chiral ${\rm SU}(2) \otimes {\rm SU}(2) $) metric, depending on two parameters. The two-parameter family of dual QFTs describes the wide class of integrable theories. Dif\/ferent reductions of these theories with respect two quantum groups describe almost all known integrable f\/ield theories with $c_{\rm UV}\leq 3$. See also recent papers~\cite{Arutyunov:2013ega,Hoare:2014pna,Klimcik:2014bta}.

\item The scattering matrix (\ref{FST}), which can be considered as non-perturbative def\/inition of SSM is related with massless scattering theory
for MSM. The second can be derived by ${\rm SL}(2)_{q}$ restriction with $q=\exp \big( i\frac{2\pi}{N+2}\big) $ of the $S$-matrix (\ref{FST}) for the value of parameter $\lambda =\frac{1}{N+2}$. Such restriction of scattering matrix of particles~(\ref{FST}) (which possesses the quantum group symmetry with $q=\exp ( i2\pi \lambda)$) to $S$-matrix of kinks is possible only for integer $N$. The similar phenomenon takes place in the sigma model with 3d target space. There also the scattering matrices for ``massive'' Ricci f\/low and massless one are related in the similar way. Probably this property is general for integrable Ricci f\/lows, which are related by simple transformation of the metric similar to (\ref{sa}),~(\ref{am}), but describe completely dif\/ferent physics.

\item The large distance pattern of sigma-models depends strongly on other possible terms in action (\ref{metr}), which are denoted as $\dots$. The f\/ields like the tachion, the Wess--Zumino--Witten term or the topological charge are usually less important in the UV behavior of theory. However, if the QFT corresponding to sigma model has broken $\mathbf{P}$, $\mathbf{T}$ symmetries (but not $\mathbf{PT}$) the action of sigma model contains the terms like $B$-f\/ield or some higher dif\/ferential forms consistent with metric. These terms appear in the Ricci f\/low equations and can be relevant in UV regime.
\end{enumerate}

\pdfbookmark[1]{References}{ref}
\LastPageEnding

\end{document}